\documentclass[prb,twocolumn,superscriptaddress]{revtex4}

\usepackage{xcolor}
\definecolor{tab20green}{rgb}{0.1725490196,0.6274509804,0.1725490196}
\definecolor{tab20red}{rgb}{0.8392156863,0.1529411765,0.1568627451}
\definecolor{tab20blue}{rgb}{0.1215686275,0.4666666667,0.7058823529}
\usepackage{amsfonts,amssymb,amsmath} 
\usepackage{graphicx}
\usepackage{dsfont}
\usepackage{soul}
\usepackage[
  colorlinks,
  linkcolor=blue,
  urlcolor=blue,
  citecolor=blue,
  plainpages=false,
  pdfpagelabels,
]{hyperref}

\graphicspath{{./figs/}}

\DeclareMathAlphabet\mathbfcal{OMS}{cmsy}{b}{n}
\newcommand{\ilfrac}[2]{\ensuremath{\,\! ^{#1} \! / _{#2}}}
\newcommand{\Z}{\mathcal{Z}}

\newcommand{\tr}{\textrm{tr}}

\newcommand{\baleqn}{\begin{equation}\begin{aligned}[b]}
\newcommand{\ealeqn}{\end{aligned}\end{equation}}
\newcommand{\baleqns}{\begin{equation*}\begin{aligned}}
\newcommand{\ealeqns}{\end{aligned}\end{equation*}}

\newcommand{\beq}{\begin{equation}}
\newcommand{\eeq}{\end{equation}}
\newcommand{\beqa}{\begin{eqnarray}}
\newcommand{\eeqa}{\end{eqnarray}}

\newcommand{\gras}[1]{\bold{#1}}

\begin{document}
\title{Thermally induced metallic phase in a gapped quantum spin liquid -- \\ a Monte Carlo study of the Kitaev model with parity projection}

\author{Chris N. Self}
\affiliation{School of Physics and Astronomy, University of Leeds, Leeds, LS2 9JT, United Kingdom}
\affiliation{Department of Physics, Imperial College London, London, SW7 2AZ, United Kingdom}
\author{Johannes Knolle}
\affiliation{Blackett Laboratory, Imperial College London, London SW7 2AZ, United Kingdom}
\author{Sofyan Iblisdir}
\affiliation{Dpt. F\'isica Qu\`antica i Astronom\'ia \& Institut de Ci\`encies del Cosmos, Facultat de F\'isica, Universitat de Barcelona, 08028 Barcelona, Spain}
\affiliation{Dpto. An\'alisis Matem\'atico y Matem\'atica Aplicada, Facultad de Matem\'aticas, Universidad Complutense de Madrid, 28040 Madrid, Spain}
\affiliation{Instituto de Ciencias Matem\'aticas, Campus Cantoblanco UAM, 28049 Madrid, Spain}
\author{Jiannis K. Pachos}
\affiliation{School of Physics and Astronomy, University of Leeds, Leeds, LS2 9JT, United Kingdom}

\date{\today}

\begin{abstract}
Thermalisation is a probabilistic process. As such, it is generally expected that when we increase the temperature of a system its classical behaviour dominates its quantum coherences. By employing the Gibbs state of a translationally invariant quantum spin liquid -- Kitaev's honeycomb lattice model -- we demonstrate that an insulating phase at $T=0$ becomes metallic purely by increasing temperature. In particular, we compute the finite temperature distribution of energies and show that it diverges logarithmically, as we move to small energies. The corresponding wave functions become critical alike as at Anderson transitions. These characteristics are obtained within a new exact Monte Carlo method that simulates the finite temperature behaviour of the Kitaev model. In particular, we take into account the projection onto the physical parity sectors, required for identifying the topological degeneracy of the model. Our work opens the possibility to detect thermal metal behaviour in spin liquid experiments.



\end{abstract}

\maketitle


\section{Introduction}

Spin liquids are commonly understood as systems where quantum fluctuations prevent magnetic ordering down to the lowest temperatures giving rise to a long-range entangled quantum phase \cite{Savary2016,knolle2018field}. An archetypical example of a spin liquid is Kitaev's honeycomb lattice model (HLM) \cite{Kitaev2006}. 
It can support Abelian anyonic phases as well as phases with non-Abelian anyons. 
Its physics is very rich and at the same time the model is analytically and numerically tractable. This tractability allowed a rigorous investigation of its dynamical correlations~\cite{Knolle2014,Knolle2015a} and thermodynamic properties~\cite{Nasu2014,Nasu2015b,Nasu2015a}.
The HLM is experimentally relevant too ~\cite{Jackeli2009}. It has been recently related to various materials such as $\text{A}_2\text{IrO}_3$ ($\text{A}=\text{Na,Li}$)~\cite{Singh2010} or $\alpha-\text{RuCl}_3$~\cite{Plumb2014,banerjee2016proximate,winter2017models,hermanns2017physics}, where the heat capacity and thermal conductivities have been measured~\cite{Hentrich2018unusual,Kasahara2018unusual}. Excitingly, a recent experimental observation~\cite{kasahara2018majorana} of a quantised thermal Hall effect in the magnetic field induced~\cite{Sears2017phase,Baek2017evidence,banerjee2018excitations} spin liquid state of $\alpha-\text{RuCl}_3$ opens the possibility for an experimental investigation of exotic phenomena related to anyonic excitations in a magnetic material. 

In the non-Abelian coupling regime the spectrum of the HLM is given in terms of fermions and vortices that bind Majorana zero modes~\cite{Kitaev2006}. While these bound Majoranas have zero energy when the vortices are well separated, they energetically split when they are brought into close proximity due to vortex-vortex couplings. Interestingly, these couplings alternate in sign depending on the distance between the vortices~\cite{Lahtinen:2011gm}. It has been demonstrated that ensembles of two-dimensional lattices of Majorana zero modes paired with random sign couplings give rise to a so called `thermal metal phase'~\cite{Altland1997,Chalker2001,Mildenberger2007,Laumann2012}. This phase, that we will refer to as `metallic', is characterised by electrically neutral energy currents traversing its bulk~\cite{Beenakker2013}. A priori, the HLM is a translationally invariant spin lattice model. Nevertheless, due to its extensive number of conserved plaquette operators, so called fluxes, it is equivalent to a sum of inhomogeneous Hamiltonians of Majorana modes that describe different sectors of the model ~\cite{Kitaev2006}. When the temperature is higher than the typical flux gap, it is natural to expect a significant contribution from Majorana Hamiltonians with a random sign of couplings from the random flux configurations. Here we address the natural question: can the HLM support metallic behaviour with extended states purely by increasing its temperature? 
We can answer this question affirmatively via our new Monte Carlo method, which is not only able to reproduce the thermodynamic properties of the model~\cite{Nasu2015a,Nasu2015b,Metavitsiadis2017}, but which also recovers the more subtle topological properties by an explicit projection onto the physical parity sectors.  

Insulating and metallic behaviours are manifestations of macroscopic quantum effects. Typically, during the transition from an insulator to a metal a redistribution of the energy eigenvalues and delocalisation of eigenstates occurs via a change of external parameters. Here we demonstrate that increasing the temperature of the translationally invariant HLM leads to a behaviour similar to a thermal metal, even if it has an insulating gap at zero temperatures. The metallic behaviour arises from Majorana fermions moving in a disordered medium of fluxes effectively created by the finite temperature.

To discriminate the characteristics of the thermal metal emerging in the non-Abelian phase we compare it to the thermal behaviour of the Abelian Toric Code phase. We begin by studying the average fermionic energy gap. The average fermion gap in the non-Abelian phase vanishes at high temperature, whereas it does not in the Abelian phase. Probing deeper, we further show that in the non-Abelian phase the distribution of fermionic energy levels diverges logarithmically at low energy and that the wave functions of these low energy states acquire a fractal character. 
These behaviours of the non-Abelian phase are similar to those of systems at the critical point of an Anderson transition between a metal and an insulator~\cite{Evers2008}. 
Our study shows that energy transport mediated by non-localised states, as in metallic systems, can be thermally activated in a spin liquid.

\section{Kitaev's honeycomb lattice model at finite temperatures} 
\label{sec:HLM}

Kitaev's honeycomb lattice model comprises spin-$1/2$ particles arranged on the vertices of a honeycomb lattice. The spins interact via the Hamiltonian
\begin{equation}
{H} = - \sum_{(i,j):\alpha} J_\alpha \sigma_\alpha^i \sigma_\alpha^j +K \sum_{(i,j,k)} \sigma_{\alpha}^i \sigma_{\beta}^j \sigma_{\gamma}^k ,
\label{eqn:Hamiltonian}
\end{equation}
where $\alpha, \beta,\gamma=x,y,z$ and $\sigma^\alpha$ are the Pauli matrices. The two-spin interaction terms are anisotropic bond-dependent Ising couplings. The three-spin terms break time-reversal invariance, with the triplets $(i,j,k)$ being consecutive indices around a plaquette.

The Hamiltonian (\ref{eqn:Hamiltonian}) has an extensive number of conserved quantities, which makes it exactly solvable. It is solved by rewriting it in terms of a $\mathbb{Z}_2$ gauge field associated with the links of the lattice, described by operators $\hat u_{ij}$, and Majorana fermions living on the vertices \cite{Kitaev2006}. The gauge field operators commute with the Hamiltonian, so we can consider fixed static choices of gauge $\gras{u} = \{{u}_{ij} =\pm 1\}$. In a given sector $\gras{u}$, the Hamiltonian is
\begin{equation}
H_{\gras{u}}= \frac{i}{4} \, \sum_{j,k=1}^{2 L^2} \, A_{jk} \, c_j c_k,
\label{eqn:ferm-Hamiltonian}
\end{equation} 
where $L$ is the linear size of the system (the number of hexagonal plaquettes), $c_j$ is a Majorana fermion living on site $j$. The couplings are given by $A_{jk}= 2 J_{jk} {u}_{jk}$ if $(j,k)$ is a nearest-neighbour pair, $A_{j,k}=2 K \sum_l {u}_{jl} \, {u}_{kl}$ if $(j,k)$ is a next-nearest-neighbour pair, otherwise $A_{j,k}=0$. As a quadratic fermionic Hamiltonian, $H_{\gras{u}}$ can be efficiently diagonalised. 
(Viewed in this way, $H_{\gras{u}}$ describes a spinless p-wave superconductor.)
The Wilson loops of the gauge field for configuration $\gras{u}$, obtained by multiplying the values of ${u}_{ij}$ around a hexagonal lattice plaquette, corresponds to a quenched configuration of vortex defects in the superconductor.
However, it is important to remember that in the original spin language these vortices will arise as quasiparticle excitations of eqn.~\eqref{eqn:Hamiltonian}. 
In the strongly dimerised limit where $J_x,J_y\ll J_z$, a perturbative expansion shows that the model behaves as the toric code (Abelian phase) \cite{Kitaev2006}, whereas for $J_x\approx J_y \approx J_z$ and $K \ne 0$, it supports Ising anyons (non-Abelian phase).

To contrast different thermal behaviours, we concentrate on the two distinct gapped phases of the model. First, we consider the Abelian (toric code) phase associated with the strongly dimerised limit, $J_x,J_y\ll J_z$ and $K=0$~\cite{Kitaev2006}. Second, we consider the non-Abelian (Ising anyon) phase, where $J_x\approx J_y \approx J_z$ and $K \ne 0$. The particle types of the toric code are the vacuum 1, the anyons $e$ and $m$ and the fermion $\epsilon= e \times m$. In the Ising anyon regime, the quasiparticle types are the vacuum $1$, the fermion $\psi$ and the Ising anyon $\sigma$. These anyons satisfy the non-Abelian fusion rule $\sigma\times\sigma=1+\psi$. Namely, if two $\sigma$ are brought together they combine to either give the vacuum, 1, or a fermion, $\psi$. In the non-Abelian regime of the HLM, vortices essentially behave as Ising anyons~\cite{Lahtinen2009}. In our Abelian case, vortices on alternating rows of plaquettes are identified with $e$ or $m$ toric code anyons~\cite{Kitaev2006}.

\subsection{Fermion parity projection}

In order to access all topologically distinct gauge field configurations we must consider the most general cases, where every gauge operator $\hat u_{ij}$ can be fixed independently to its eigenvalues $u_{ij} =\pm 1$.
However, if we allow these completely general gauge configurations we must follow the Kitaev fermionisation~\cite{Kitaev2006}. 
This approach has the complication that the fermionic Hamiltonian eqn.~\eqref{eqn:ferm-Hamiltonian} lives in an extended Hilbert space.
Given an eigenstate $|\Psi_u\rangle$ of some $H_{\gras{u}}$, it is necessary to apply a projection operator $\mathcal{P}$ in order to get an eigenstate of $H$, eqn.~\eqref{eqn:Hamiltonian}. This projector is made of two parts: $\mathcal{P}=\mathcal{S} \mathcal{P}_0$~\cite{Pedrocchi2011}. The part $\mathcal{S}$ acts purely on the gauge degrees of freedom $u_{ij}$, in such a way that it does not mix topologically distinct gauge configurations. Whereas $\mathcal{P}_0$ acts on the fermions, in each sector $\gras{u}$ it is given by
\begin{equation*}
\mathcal{P}_0 = \frac{1}{2} \left( 1 + \text{det}(Q) \bigg( \prod_{(i, j)} u_{ij} \bigg) \mathcal{F} \right) \, .
\label{eqn:ferm-proj-op}
\end{equation*}
The product over pairs $(i, j)$ runs over every lattice link and must be consistent with the chosen orientations of the links~\cite{Pedrocchi2011}. The orthogonal matrix $Q$ is the matrix that brings $H_{\gras{u}}$ into Majorana normal form (see section 3 of Kitaev 2006~\cite{Kitaev2006}). The operator $\mathcal{F}$ counts the parity of the occupancies of the diagonal fermions $f_p$, $f^\dagger_p$, given by
\begin{equation*}
\mathcal{F} = \prod_p (1 - 2 f^\dagger_p f_p)
\end{equation*} 
The eigenvalues of $\mathcal{F}$ are +1 (when an even number of modes are excited) and -1 (when the number is odd). The operator $\mathcal{P}_0$ therefore projects into a particular parity of the diagonal fermions, fixed by the configuration $\gras{u}$.

Note that previous studies of the HLM at non-zero temperature employed alternative fermionisation approaches~\cite{Nasu2015a,Nasu2015b,Metavitsiadis2017}. These have the advantage that the states obtained in that way do not need to be projected. However, these fermionisations correspond to only considering certain restricted sets of gauge configurations. Under these restrictions it is not possible to properly probe the topological degeneracy of the ground state.
By implementing the projection, we can study the finite temperature behaviour of the fermion parity and probe the contribution of the topological ground state degeneracy to the low temperature thermal entropy, as seen in the following.

\begin{figure*}
\centering
\includegraphics[width=0.9\textwidth]{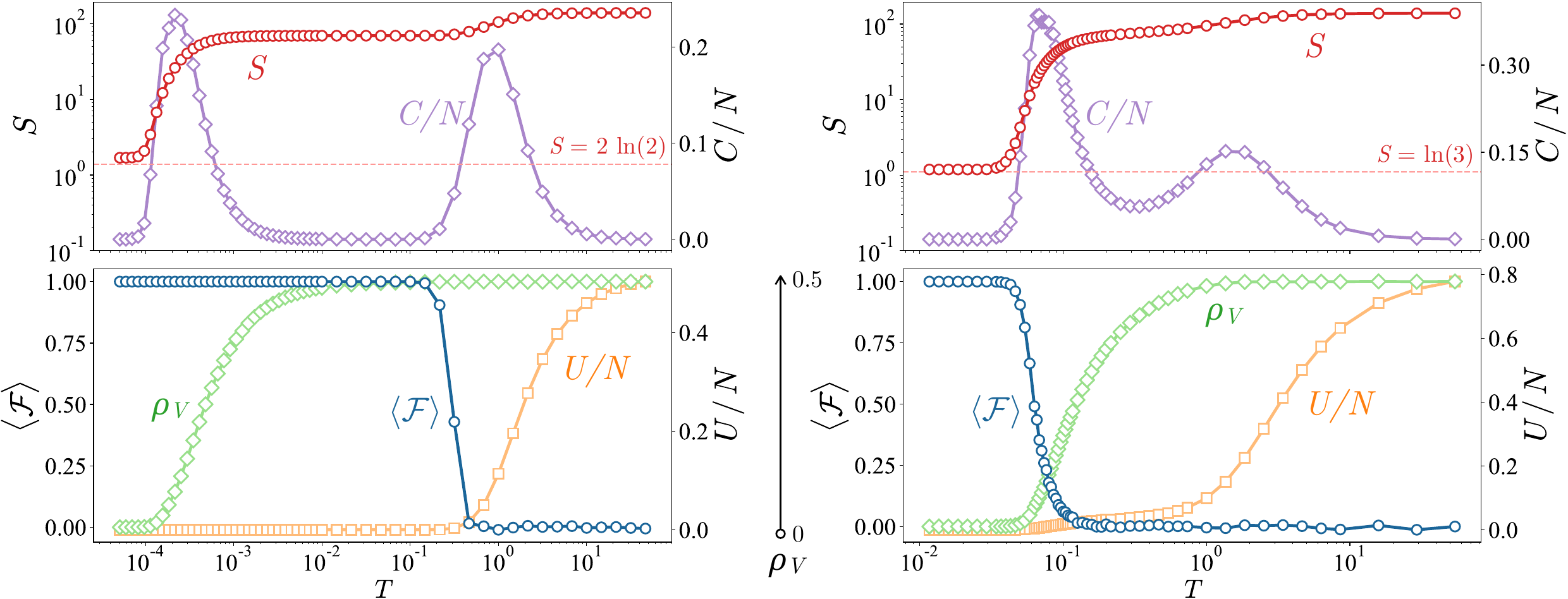}
\caption{
Thermodynamic quantities in the (left) Abelian and (right) non-Abelian regime. 
In the top panel we plot the entropy $S$ (red circles) and the specific heat $C/N$ (purple rhombus).
The distinctive two peak structure of $C/N$ is clearly visible, with the lower temperature peak associated with the activation of the vortex excitations and the higher temperature peak with the activation of the fermionic excitations. At low temperatures $S$ identifies the different topological degeneracies of the ground states, which is fourfold in the Abelian case ($S =2\ln2$) and threefold in the non-Abelian case ($S=\ln 3$). At higher temperatures the entropic signature of topological order is lost due to the proliferation of vortices. The lower panel plots vortex density $\rho_V$ (green rhombus), internal energy per particle $U/N$ (orange squares) and average fermion parity $\langle \mathcal{F} \rangle$ (blue circles). The vortex density $\rho_V$ is zero at low temperature, and increases to $1/2$ when the first peak of $C/N$ occurs, while the internal energy becomes non-zero at the second peak. The average fermion parity $\langle \mathcal{F} \rangle$ shows distinctly different behaviours in the Abelian and non-Abelian regimes for intermediate temperatures between the two peaks.
Data is shown for $L=10$ system size, with couplings $J_x = J_z = 0.25$, $J_z=1$, $K=0$ in the Abelian case and $J_x = J_y = J_z=1$, $K=0.1$ in the non-Abelian case.
}
\label{fig:thermo_and_ferm_parity}
\end{figure*}

\subsection{Finite temperature expectation values}
\label{sec:HLM:expt-vals}

The finite temperature physics of Hamiltonian (\ref{eqn:Hamiltonian}) can be numerically studied by randomly sampling different gauge configurations $\gras{u}$, in each case diagonalising the fermionic Hamiltonian $H_{\gras{u}}$ fully. This is possible since the partition function at temperature $T = 1/\beta$ can be written
\begin{gather*}
\Z(\beta) = \tr(e^{-\beta H}) = \sum_\gras{u} {\Z}'(\beta;\gras{u})  \, ,
\label{eqn:full-part-func}
\end{gather*}
where ${\Z}'(\beta;\gras{u}) \equiv \tr'( \mathcal{P} \,  e^{-\beta H_\gras{u}} )$ with $\tr'$ is over the fermionic degrees of freedom only. The projected fermionic partition functions ${\Z}'(\beta;\gras{u})$ can be calculated exactly~\cite{chrisUpcoming}. The weighting of each $\gras{u}$ in the thermal state is thus given by $p_\beta(\gras{u}) = {{\Z}'(\beta;\gras{u})}/{\Z(\beta)}$, and finite temperature expectation values for observable $\mathcal{O}$ can be estimated by approximating
\begin{align*}
\langle \mathcal{O} (\beta) \rangle &= \sum_{\gras{u}} p_\beta(\gras{u}) \, \frac{\tr'( \mathcal{P} \, e^{-\beta H_\gras{u}} \mathcal{O}_\gras{u} )}{\overline{\Z}(\beta;\gras{u})} \nonumber \\ 
&\equiv \sum_{\gras{u}} p_\beta(\gras{u}) \, \langle \mathcal{O}_\gras{u} (\beta) \rangle' \, ,
\end{align*}
using Monte Carlo methods.
For all quantities of interest the projected fermionic expectation values $\langle \mathcal{O}_\gras{u} (\beta) \rangle'$ can again be computed exactly~\cite{chrisUpcoming}. The heat capacity of the full model is given by
\begin{equation*}
C = \beta^2 \, \text{Var}_\gras{u} \Big( \langle H_\gras{u} \rangle' \Big) + \text{Mean}_\gras{u}\Big( \beta^2 \big\langle \, (H_\gras{u} - \langle H_\gras{u} \rangle')^2 \, \big\rangle' \Big) \, ,
\end{equation*}
where $\text{Var}_\gras{u}$ and $\text{Mean}_\gras{u}$ are calculated with respect to the probability distribution $p_\beta(\gras{u})$. Numerically these variances, $\text{Var}_\gras{u}$, are estimated along with their uncertainties using bootstrap resampling~\cite{Efron1979}. At infinite temperature the state consists of $N$ decoupled spin-1/2 particles, which has thermal entropy $S = N \ln(2)$. The thermal entropy can then be obtained by computing
\begin{equation*}
S(T) = N \ln(2) - \int_T^\infty \text{d}T' \, \frac{C(T')}{T'} \, .
\end{equation*}

\subsection{Numerical results} 

To analyse the thermal behaviour of the HLM we calculate the specific heat, \emph{i.e.} the heat capacity per spin, $C / N$, thermal entropy $S$, vortex density, $\rho_V$, internal energy per spin, $U/N$ ($U = \langle H \rangle$) and average fermion parity $\langle {\cal F}\rangle$ of the model. These are plotted as functions of temperature for both the toric code and Ising coupling regimes in Fig.~\ref{fig:thermo_and_ferm_parity}. 

First, we see that the splitting of the spins into gauge and fermionic degrees of freedom is not only a mathematical trick for solving Kitaev's honeycomb lattice model. The model exhibits a physical fractionalisation of its spins into gauge vortices and fermions. We observe the distinctive two peak structure of the specific heat, indicating the fractionalisation of the spins, seen in previous studies~\cite{Nasu2015b}. For both the toric code and Ising regime, the low $T$ peak is associated with a change of the vortex density, in which it rises from its ground state value $\rho_V = 0$ towards a completely disordered vortex configuration $\rho_V = 1/2$. The higher temperature peak is linked to a large change in internal energy $U$, corresponding to an unfreezing of the fermionic modes. These peaks are well separated by several orders of magnitude of $T$ in the toric code limit, whereas they are closer and blur together to a certain extent in the Ising regime. Nevertheless, the specific heat in both regimes exhibit similar qualitative behaviour. 

Second, in Fig.~\ref{fig:thermo_and_ferm_parity} (upper panel) we plot the behaviour of the thermal entropy, $S$, obtained by integrating the heat capacity down from infinite temperature. At small temperatures the thermal entropy is able to identify the topological degeneracy of the ground state. Specifically, the Boltzmann formula gives $S = \ln(\Omega) \,$, for $\Omega$ the number of microstates (with the Boltzmann constant set equal to one). As $T \rightarrow 0$, $\Omega$ gives the degeneracy of the ground state whose correct value can only be obtained by employing the projection onto physical states. Hence, the thermal entropy is in excellent agreement with the topological degeneracies expected for the toric code anyons $S=2\ln2$ and for the Ising anyons $S=\ln 3$. However, for larger temperatures when the vortices start to proliferate the entropic signatures of topological order are washed out~\cite{Iblisdir2009scaling,iblisdir2010thermal}. 

In contrast to the thermal entropy, the fermionic parity $\langle \mathcal{F} \rangle$ shows interesting differences between the two regimes at higher temperatures and in fact hints at even more complex differences between the two. We begin by describing the behaviour we would expect $\langle \mathcal{F} \rangle$ to have based purely on the anyon models.
The $T \rightarrow 0$ state always has no vortices. In the toric code case this already rules out the possibility of odd fermion parity states because of the fusion rules. States containing two fermionic quasiparticles would be allowed, however these have very high energy. Whereas in the Ising regime odd fermion parity states that do not contain any vortices are allowed by the fusion rules (they correspond to states with anyonic flux around both handles of the torus). However, since they contain a fermion they have higher energy, so these states do not occur at low temperature. Hence at low $T$ we expect to find $\langle \mathcal{F} \rangle=1$ in both regimes. In contrast, at high temperatures we would expect different values of $\langle \mathcal{F} \rangle$ in the two cases. As $T \rightarrow \infty$ the vortices becomes completely disordered, $\rho_V = 1/2$. In the toric code case half of these configurations will contain odd numbers of $e$ (and $m$) anyons and half even numbers. From the fusion rules these correspond to odd and even fermion parities respectively, and averaging over them gives $\langle \mathcal{F} \rangle = 0$. Whereas in the Ising theory the only way to generate a state with total odd fermion parity is to add non-trivial anyon flux around both torus handles. This is one of four flux configurations and so we would instead expect to find $\langle \mathcal{F} \rangle = \ilfrac{3}{4}(+1) + \ilfrac{1}{4}(-1) = 0.5$. Fig.~\ref{fig:thermo_and_ferm_parity} (lower panel) plots the fermion parity as a function of temperature. We see that at low temperature both regimes agree with the predictions of the anyon models. At high temperature, however, only the toric code case agrees with our predictions. The fermion parity also becomes zero in the Ising case at high temperatures. This is a first indication that the thermal proliferation of Ising vortices has driven a transition out of the topologically ordered phase (described by the anyon model) to a new kind of phase. We will demonstrate in the following that this phase is a thermal metal.

\section{Thermal metal behaviour} 

Of central interest to us is the behaviour of a two-dimensional Majorana lattice with random sign couplings. Such random configurations are renormalisation fixed points. Indeed, if block renormalisation is employed the resulting model is again a Majorana lattice with random sign of couplings~\cite{Evers2008}. These signs can be interpreted as a $\mathbb{Z}_2$ gauge field that gives rise to Majorana binding vortices. It has been shown that a Majorana lattice with $\mathbb{Z}_2$ field couplings can be faithfully modelled by the same system with homogeneous couplings, superposed by an independent lattice of Majoranas at the position of the vortices created by the random signs~\cite{Lahtinen2012}. As the vortices are farther apart than the original lattice spacing they interact more weakly. So this vortex lattice contributes states with energies below the band gap of the original lattice model. Moreover, the sign of the vortex couplings is random due to the random position of the vortices. This picture holds recursively providing subgap states all the way to zero energy. As a result we expect the randomisation of the coupling signs to give rise to a gapless phase. Moreover, as randomness persists with renormalisation, this critical phase should be described by an Anderson transition with wave functions exhibiting multifractality~\cite{Evers2008}. The divergence of the number of fermionic states with close-to zero energy and the critical multifractal behaviour of the wave functions are the defining characteristic of a thermal metal phase.

Two-dimensional Majorana lattice with random sign couplings have been numerically proven to support  thermal metal phase~\cite{Laumann2012,Lahtinen2014}. Thermal metal behaviour has also been seen in topological superconductors at temperatures above the Kosterlitz-Thouless transition point where vortices that bind Majorana zero-modes proliferate ~\cite{Bauer2013}. However, in both of these examples the vortices are introduced by hand as they are not part of the thermal states of the system. Hence, ensembles of different realisations of the system have to be involved.

In the HLM case, in contrast, the vortices naturally occur as finite temperature excitations, as shown in Fig.~\ref{fig:thermo_and_ferm_parity}. To intuitively understand why such a thermal metal phase can emerge in this situation, consider the system at $T=0$ in the presence of various patterns of vortices. When vortex quasiparticles of the non-Abelian regime are brought into proximity, they couple with a sign that depends on their relative position~\cite{Lahtinen2012}. As temperature increases, in the HLM the average vortex density increases to the point that the vortex-vortex couplings become important. Moreover, the random distances between vortices mean the pairing energies between the vortex bound states are positive or negative at random. Thus we also expect a thermal metal phase to arise in the HLM at finite temperature. 

\subsection{Fermionic gap}

\begin{figure}[t]
\centering
\includegraphics[width=\columnwidth]{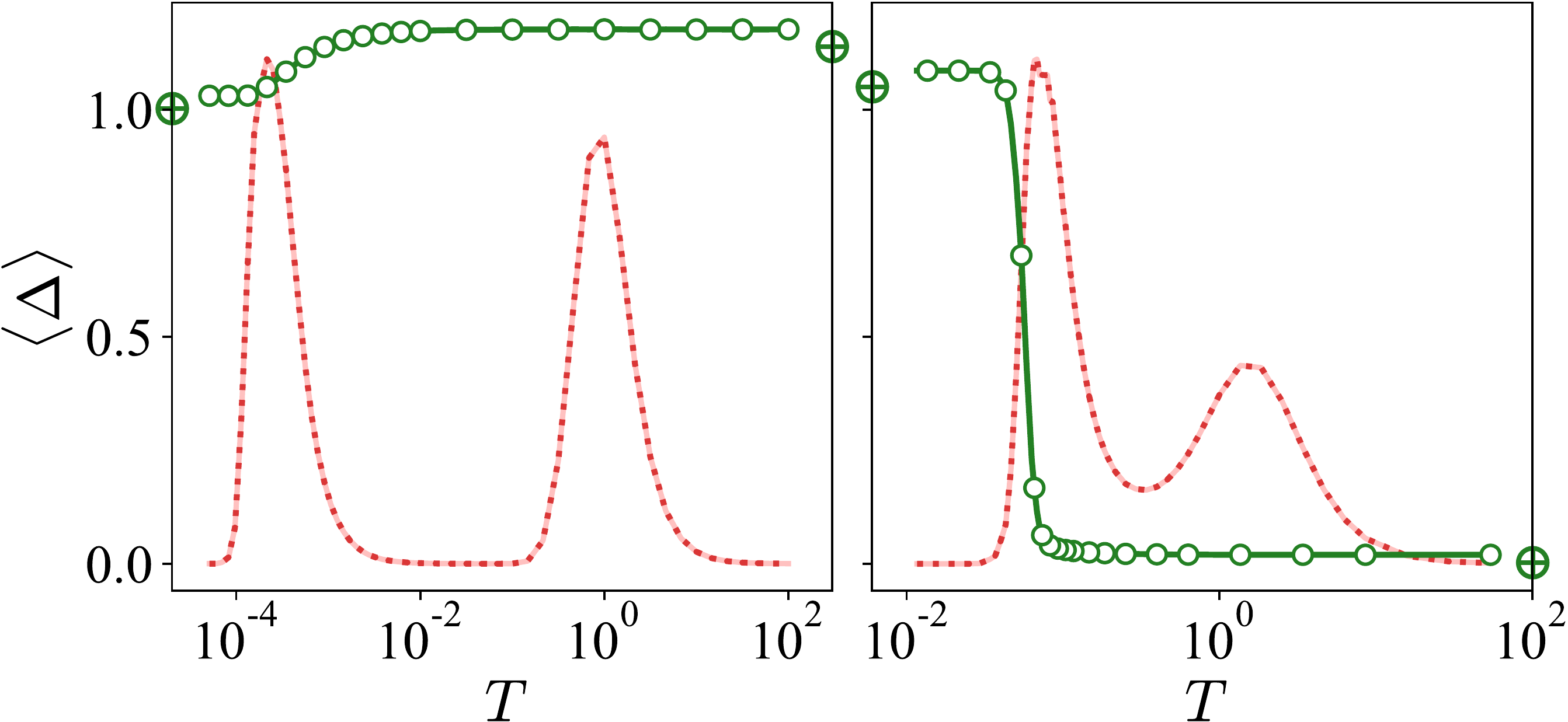}
\caption{
Average fermion gap for the (left) Abelian and (right) non-Abelian cases, with the specific heat profile superimposed.
The fermions in the Abelian limit are gapped at all temperatures, whereas in the non-Abelian case they become gapless when vortices start proliferating.
The large system size limits of the fermion gap for $T\rightarrow0$ and $T\rightarrow\infty$ (obtained in appendix~\ref{app:finite-size-gap}) are indicated with markers on the axes. Data is shown for $L=10$ system size with couplings set to $J_x = J_z = 0.25$, $J_z=1$, $K=0$ in the Abelian case and $J_x = J_y = J_z=1$, $K=0.1$ in the non-Abelian case.
}
\label{fig:fermion-gap}
\end{figure}

Thermal metals conduct heat through their fermionic degrees of freedom. This metallic conduction property requires a vanishing fermion gap. The behaviours of the average fermion gap, $\langle \Delta \rangle$, in the Abelian and non-Abelian regimes are shown in Fig.~\ref{fig:fermion-gap}. In these plots the fermionic gap is averaged over the vortex sectors that are thermally excited for a given temperature $T$. We see that the Abelian regime remains gapped at all temperatures. 
Whereas, in the non-Abelian case the average fermionic gap vanishes above the temperature at which vortices begin to proliferate. In appendix~\ref{app:finite-size-gap} we address the finite size scaling of the fermion gap in the zero temperature and infinite temperature limit. We give evidence that in the non-Abelian phase the infinite temperature limit $\langle \Delta \rangle$ tends to zero as $L \rightarrow \infty$.



A vanishing fermion gap is a necessary condition for a thermal metal. However, since the non-Abelian vortices bind Majorana modes at zero-energy the vanishing of the gap is not sufficient. It cannot distinguish between a metallic phase (which has fermionic energy levels occuring continuously down to zero energy) and an insulating phase with additional zero energy localised states associated with the vortices (isolated subgap fermionic states at zero energy). 
Further, a thermal metal phase is expected to display its own distinct behaviours in both the distribution of energy levels at low energy and the spatial properties of the wave functions~\cite{Altland1997,Mildenberger2007}. In the following we probe both of these properties in detail.

\subsection{Density of states}

\begin{figure}[t]
\centering
\includegraphics[width=\columnwidth]{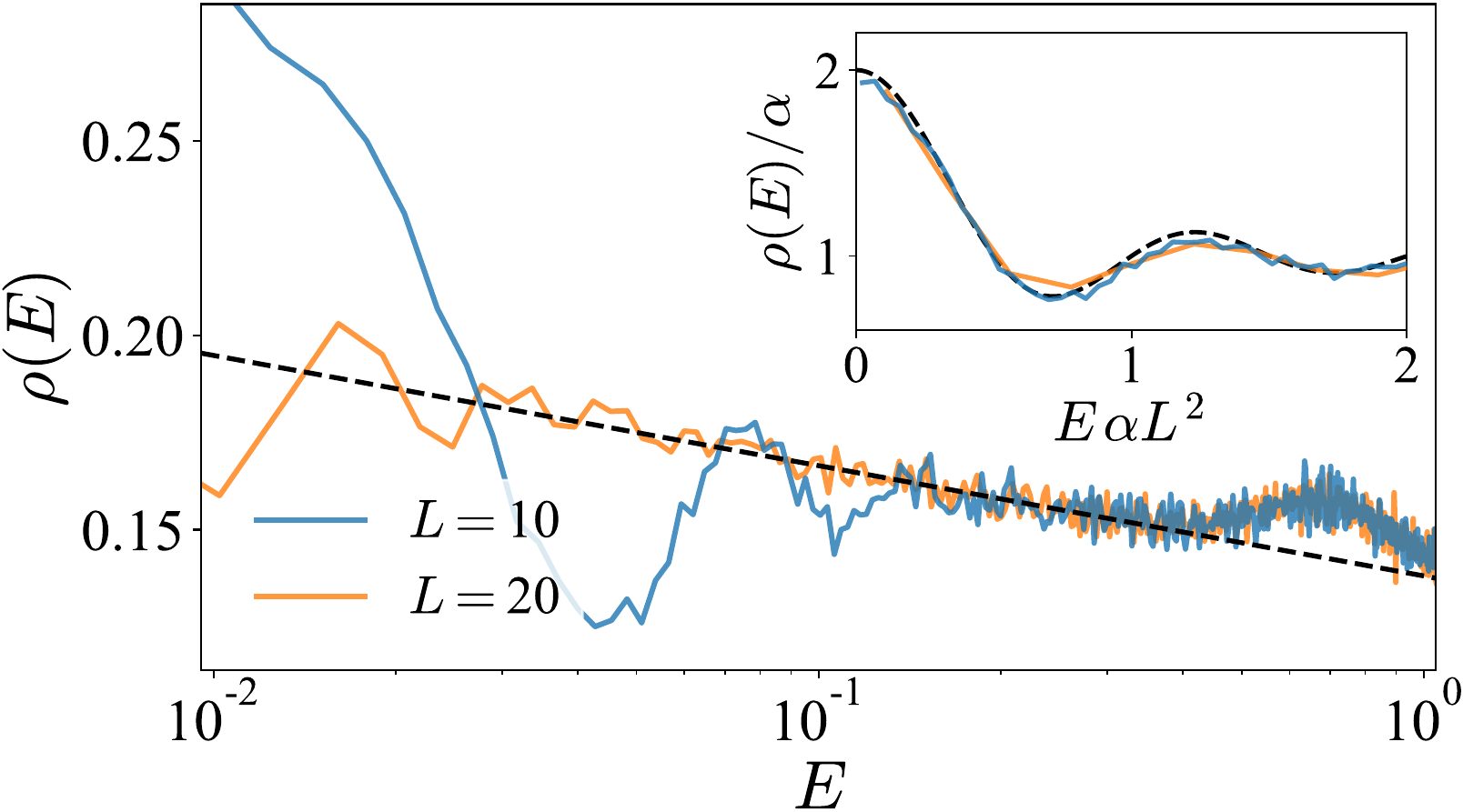}
\caption[DOS]{Distribution of fermion energies $\rho(E)$ for low $E$ in the non-Abelian case. Computed for system sizes $L=10$ and 20 at fixed temperature $T \approx 0.4$.
At low energies $\rho(E)$ diverges logarithmically with decreasing $E$. This is indicated by the black dashed line, which is a fit to the $L=20$ data.
For the smallest values of $E$ oscillations are visible. (Inset) The oscillations are collapsed onto curves of the form Eqn.~\eqref{eqn:rmt-osc}.
The system couplings are set to $J_x = J_y = J_z=1$, $K=0.1$.}
\label{fig:DOS}
\end{figure}

The fermionic distribution of energies (DOE) is given by $\rho(E) = \langle \, \sum_n \delta(E - \varepsilon_n) \, \rangle$, where the $\varepsilon_n$ are the fermionic energies in a single vortex sector and the average is taken over the vortex sectors visited at temperature $T$~\endnote{Viewing the system as an ensemble of disordered fermionic Hamiltonians with quenched vortex disorder this would be the ensemble averaged density of states. This is the situation in which thermal metals are usually discussed. However, since the vortex disorder we consider arises from the thermal state of the Kitaev spin Hamiltonian, we avoid that term so as not to confuse $\rho(E)$ with the density of states of that underlying Hamiltonian.}. In a thermal metal, $\rho(E)$ diverges logarithmically with decreasing energy at low energies~\cite{Mildenberger2007,Bocquet:1999bb}. To clearly see this feature of $\rho(E)$ we find it necessary to go to larger systems than are accessible when using the algorithm that properly samples from the thermal state, e.g. our $L=10$ data. For system size $L = 20$ data is obtained by randomly sampling over vortex sectors, matching the average vortex density to the value obtained from smaller system sizes (see Appendix~\ref{app:sim:vortices}). The logarithmic divergence is extracted by a fit to $L=20$, and is in agreement with the behaviour of $L=10$, as shown in Fig.~\ref{fig:DOS}. For the very lowest energies, $\rho(E)$ additionally exhibits characteristic oscillations predicted by random matrix theory~\cite{Bocquet:1999bb,wigner1993class}. In this limit the DOE is given by
\begin{equation}
\rho(E) = \alpha \left( 1 + \frac{\sin(2\pi E \alpha L^2)}{2\pi E \alpha L^2} \right) \, .
\label{eqn:rmt-osc}
\end{equation}
The parameter $\alpha$ can be obtained by numerically solving the consistency  equation $\rho(\ilfrac{1}{\alpha L^2}) = \alpha$ and $\alpha$ is expected to diverge logarithmically with $L$~\cite{Mildenberger2007}. The first few oscillation periods of the random matrix theory are shown in Fig.~\ref{fig:DOS} (Inset). Hence, the energy distribution is in agreement with the thermal metal phase.

\subsection{Weak multifractality of wavefunctions}

\begin{figure}[t]
\centering
\includegraphics[width=\columnwidth]{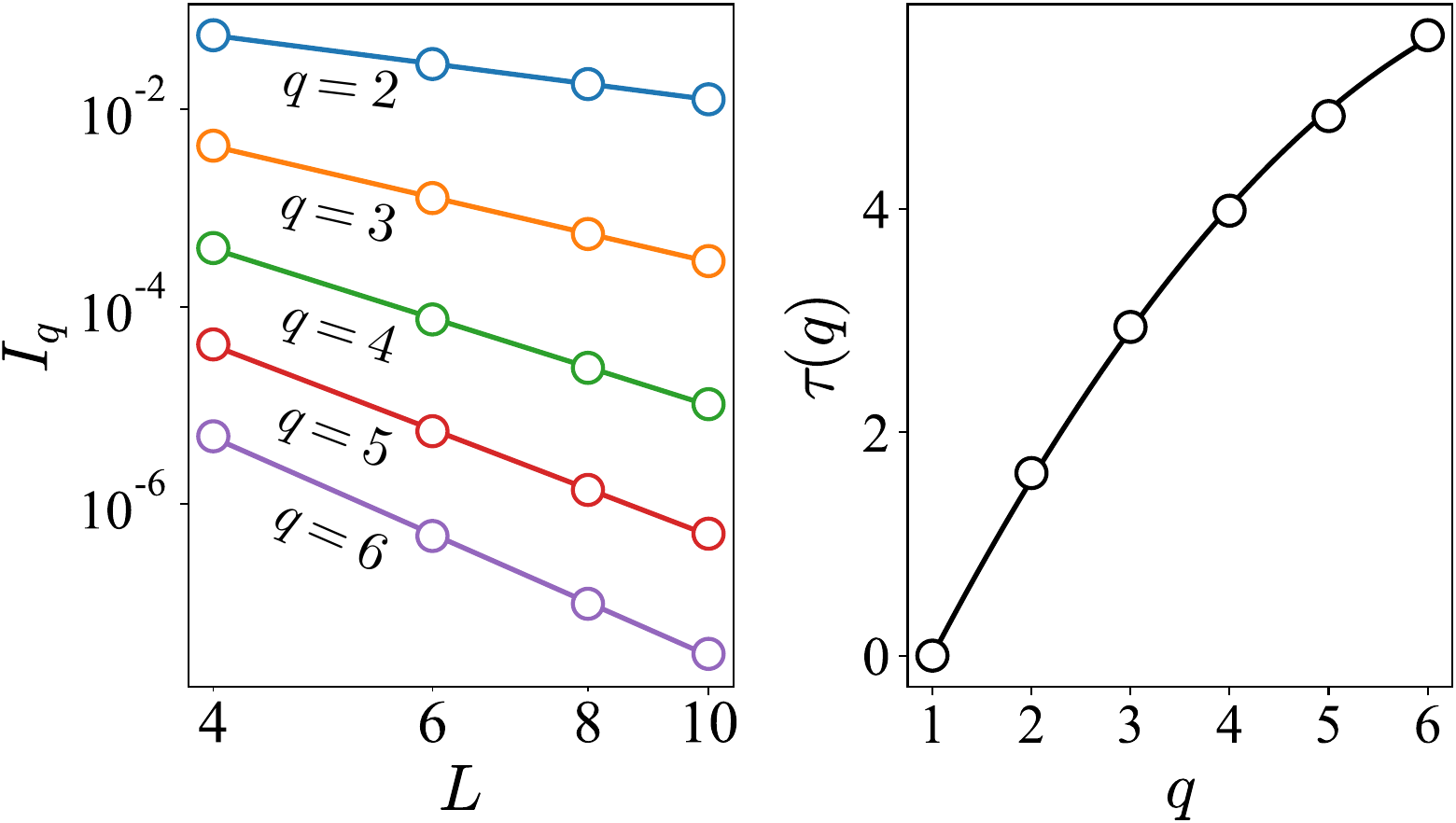}
\caption[IPR]{Weak multifractality of the fermionic wave functions in the non-Abelian case at high temperatures. (Left) Inverse participation ratios $I_q$ of the lowest-lying fermionic eigenmode as a function of system size $L$ (on a log-scale). The $I_q$ are averaged over the thermal distribution of vortices at $T \approx 0.4$. (Right) Exponents $\tau(q)$ (defined in Eq.(\ref{eq:multifractality})) as a function of $q$ demonstrating multifractal critical behaviour, in-between the metallic $\tau(q)=2(q-1)$ and the insulating $\tau(q)=0$ behaviours.}
\label{fig:IPR}
\end{figure}

Beyond the energy eigenvalues, the wave functions of thermal metals exhibit \emph{weak multifractality}, which means they are spatially extended in a particular way~\cite{Falko:2007bl}. This behaviour is shared with systems that exhibit Anderson localisation, where weak multifractality occurs at the critical point between insulating and conducting phases~\cite{Evers2008}.
Whether states are localised or extended can be studied via the inverse participation ratios of the fermionic eigenfunctions $\psi_n(\mathbf{r})$~\cite{Evers2008}, given by
\begin{equation}\label{eq:multifractality}
I_q = \int \textrm{d}^2 r \, |\psi_n(\mathbf{r})|^{2q} \sim \frac{1}{L^{\tau(q)}} .
\end{equation}
The exponent is often written $\tau(q) = (q-1) D_q$ where $D_q$ is called the \emph{fractal dimension}.
For pure metallic states $D_q$ is equal to the spatial dimension, $D_q = 2$, and for entirely localised states $D_q = 0$. 
However, weak multifractality implies $D_q$ becomes dependent on $q$, and behaves as $D_q = 2 - \gamma q$~\cite{Falko:2007bl}. Fig.~\ref{fig:IPR} shows the exponent $\tau(q)$ plotted as a function of $q$ for the lowest-lying fermionic state, averaged over vortex sectors at temperature $T \approx 0.4$. Fitting to the data we obtain the fractal dimension $D_q = 1.7(8) - 0.12(3) q$. Hence, the eigenstates also exhibit the critical Anderson behaviour expected from a thermal metal phase. 

\section{Conclusions} 

We have analysed the thermal behaviour of the Abelian and non-Abelian quantum spin liquid phases of the Kitaev honeycomb model. Using Monte Carlo simulations, we have studied the thermal distributions of vortices, finding thermodynamic behaviours that agree with past studies~\cite{Nasu2015a,Nasu2015b}. Importantly, we find that the HLM at $T>0$ enters a thermal metal phase, with logarithmic divergent distribution of energies and with wave functions that exhibit multifractality. 

Compared to the study of thermal metals in disordered integer quantum Hall effect~\cite{Laumann2012} that required fine tuning to zero chemical potential, the metallic phase obtained here is stable against changes in the coupling constants that do not move the system out of its non-Abelian phase. Compared to the p-wave superconductor~\cite{Bauer2013} the metallic phase appears without having to introduce vortices by hand; once in the non-Abelian phase, the only necessary knob is the temperature. Also, no ensemble averaging is necessary to get different patterns of vortices; throughout, we have only needed to deal  with the Gibbs state of a single translationally invariant Hamiltonian. We believe this distinction is important and could eventually contribute to the experimental observation of a thermal metallic phase, \emph{e.g.} in magnetic field-tuned $\alpha$-RuCl$_3$ above the temperature window which has recently shown signatures of a quantised thermal Hall effect~\cite{kasahara2018majorana}.

Our work paves the way for a number of future investigations: It will be important to analyse how the signatures of the thermal metal become manifest in experimental observables, e.g. the temperature or system size scaling of the thermal conductivity or the frequency dependence of the low energy Raman response \cite{Sandilands2015scattering,Brent2016}. A challenging question will be to investigate whether the metallic phase survives the addition of integrability breaking terms that render static fluxes dynamical or the coupling to acoustic phonons inevitably present in real materials. Finally, beyond the quantum spin liquid context we expect that temperature induced localisation-delocalisation transitions could also emerge in certain lattice gauge theories \cite{Smith2017,Prosko2017,smith2018dynamical} recently discussed in the context of non-ergodic phases. 

\begin{acknowledgments}

We thank Chris Turner for helpful discussions while carrying out this work. This work was undertaken on ARC2, part of the High Performance Computing facilities at the University of Leeds, UK. SI acknowledges support from MINECO (grant MTM2014-54240-P), Comunidad de Madrid (grant QUITEMAD+CM, ref. S2013/ICE-2801), and Severo Ochoa project SEV-2015-556. This project has received funding from the European Research Council (ERC) under the European Union's Horizon 2020 research and innovation programme (grant agreement No 648913). CS and JKP were supported by the EPSRC grant EP/R020612/1.

\end{acknowledgments}


\appendix

\section{Monte Carlo procedures}

In this appendix we give more precise descriptions of the Monte Carlo procedures employed in the main text.

\subsection{Sampling gauge configurations}
\label{app:sim:gauge}

\begin{figure}[t]
\centering
\includegraphics[width=\columnwidth]{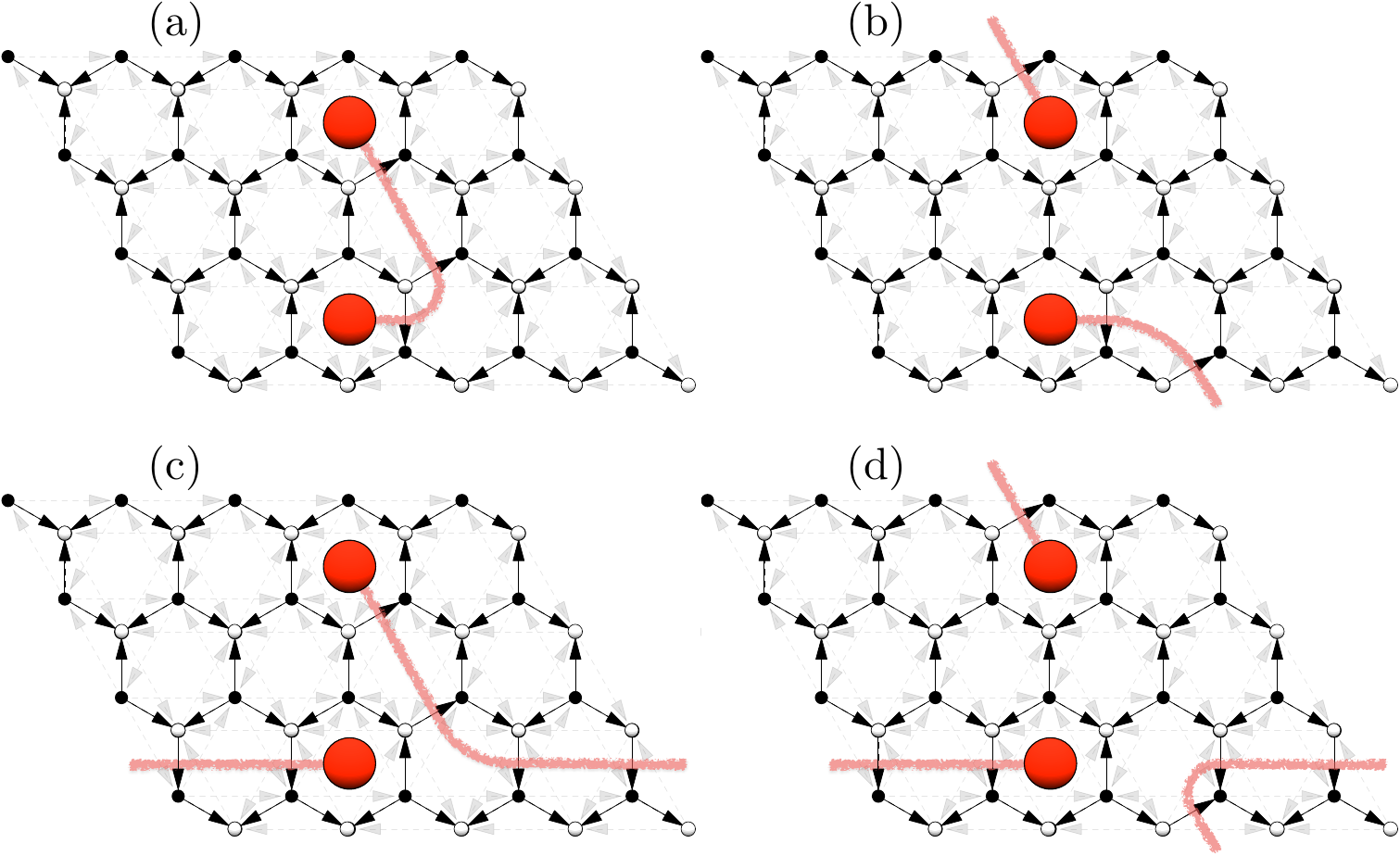}
\caption[Illustration of the Markov step proposal used in our numerics]{Markov step proposal. (a) two plaquettes are selected at random and an arbitrary rectangular path of flipped links is drawn between them. With equal probabilities, this gauge string is combined with (a) nothing, (b) a non-trivial loop wrapping the lattice up-down, (c) a non-trivial loop wrapping the lattice left-right, or (d) both of these non-trivial loops.}
\label{fig:large-mc-step}
\end{figure}

The expressions given in section~\ref{sec:HLM:expt-vals} can be approximated using Monte Carlo sampling. The aim is to choose $M \sim 50,000$ gauge configurations such that the probability of any $\gras{u}$ being chosen is ${{\Z}'(\beta;\gras{u})}/{\Z(\beta)}$. We achieve this with a Markov chain. Beginning with the uniform $u_{i,j} = +1$ gauge configuration (which is in the no-vortex sector), at each step a new gauge configuration $\gras{u}_\text{prop}$ is proposed. The proposal will be described in detail below. Each step the current $\gras{u}_\text{curr}$ is replaced by the proposed $\gras{u}_\text{prop}$ with probability:
\begin{equation*}
\min\left[ \, 1, \, \frac{{\Z}'(\beta;\gras{u}_\text{prop})}{\Z(\beta)} \! \bigg/ \frac{{\Z}'(\beta;\gras{u}_\text{curr})}{\Z(\beta)} = \frac{{\Z}'(\beta;\gras{u}_\text{prop})}{{\Z}'(\beta;\gras{u}_\text{curr})} \, \right] \, .
\label{eqn:HC-metropolis-rule}
\end{equation*}
After a large number of attempted replacements ($\sim L^2$) we add the current gauge configuration to the sample, this timescale we refer to as a \emph{sweep}. This continues until we have collected enough configurations. Each Markov step is relatively slow as computing the acceptance probability requires rediagonalising the fermionic Hamiltonian. We additionally use parallel tempering to help speed up the convergence of statistical estimates~\cite{Earl2005}. Autocorrelation effects in the Markov chain are quantified by estimating the integrated autocorrelation time and this is then included in the quoted statistical uncertainties~\cite{Geyer1992}.

The proposed change $\gras{u}_\text{curr} \rightarrow \gras{u}_\text{prop}$ is generated by selecting two random plaquettes and drawing a string of gauge flips between them, figure \ref{fig:large-mc-step}(a). This might select the same plaquette twice, in which case the string is empty. The string is combined with a randomly selected topologically non-trivial loop of gauge flips, as shown in figures \ref{fig:large-mc-step}(a)-(d). This choice of proposal effectively ignores the trivial gauge loops but otherwise is sampling over all vortex sectors and all non-trivially different gauge configurations.

\subsection{Sampling of vortex sectors}
\label{app:sim:vortices}

A faster sampling procedure is to sample from random vortex patterns with fixed densities. This allows us to reach much larger system sizes than are accessible when sampling over gauge configurations. By matching these fixed densities to the known thermal density of vortices (for example the profile of $\rho_V$ obtained from the full simulation of small system sizes, e.g. Fig.~\ref{fig:thermo_and_ferm_parity}) this captures most of the thermal physics in a more efficient way. For high temperatures this has been shown to work well~\cite{Metavitsiadis2017,Nasu2015b}. However, we do not expect this approximation to perfectly capture the physics around the first peak in specific heat. Since the vortices are interacting~\cite{Lahtinen:2011gm}, the thermal patterns of vortices at lower densities are expected to be quite different from purely random patterns. Hence we cannot use this approach to probe the onset of the thermal metal phase.

\section{Finite size scaling of the fermion gap}
\label{app:finite-size-gap}

It is simple to sample from the zero temperature and infinite temperature gauge configurations. In this appendix we study the fermion gap in these limits and address their scaling with system size. As an additional check on our Monte Carlo results we compare the temperature extremes of our Monte Carlo data to these limiting cases. 

\begin{figure}[t]
\centering
\vspace{10pt}
\includegraphics[width=\columnwidth]{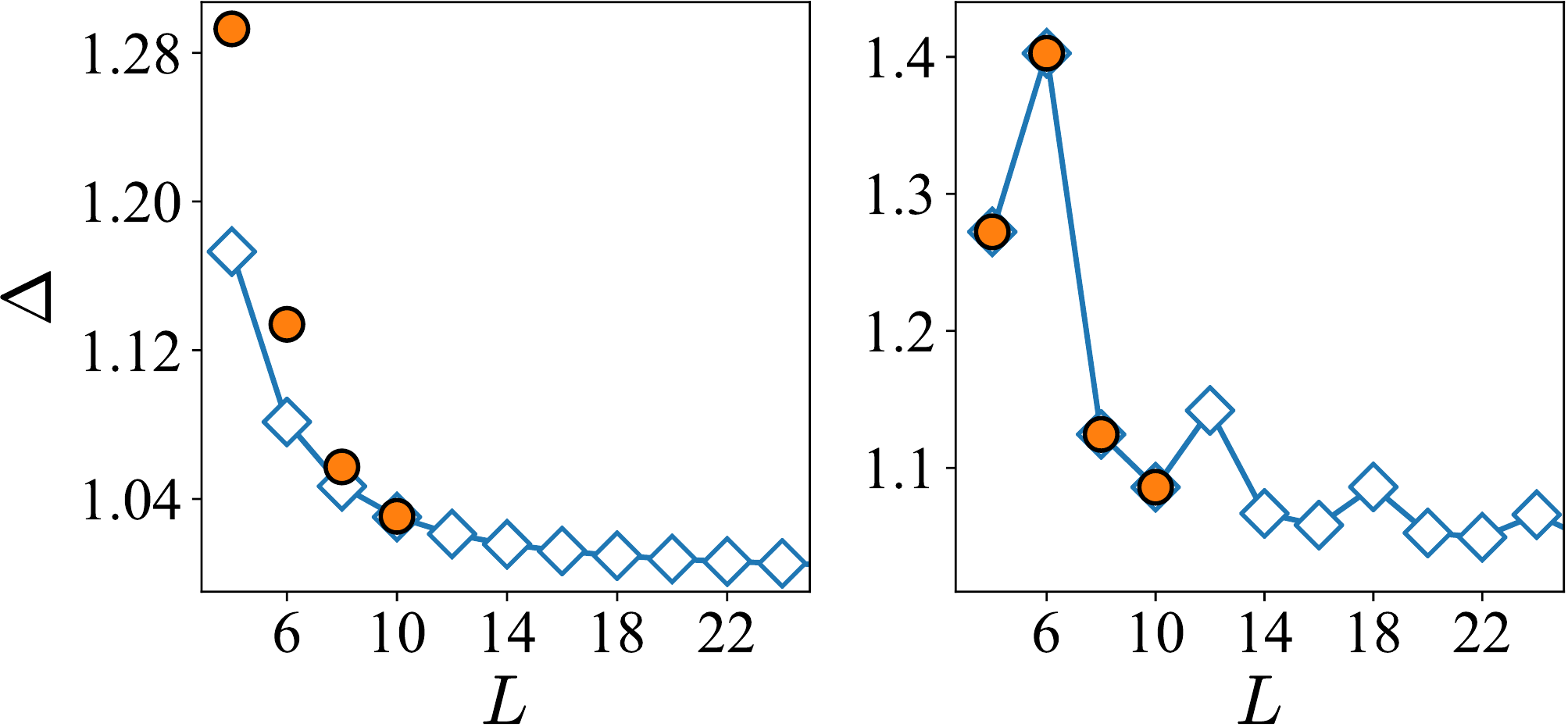}
\caption[smallT_gap]{Zero temperauture fermion gap as a function of system size for (left) Abelian, $J_x = J_z = 0.25$, $J_z=1$, $K=0$, and (right) non-Abelian, $J_x = J_y = J_z=1$, $K=0.1$, couplings. Both cases quickly decay towards a fixed value. The fermion gaps obtained from the lowest temperature Monte Carlo simulations are displaying (orange circles) on the plots.}
\label{fig:smallT_gap}
\end{figure}

\begin{figure}[t]
\centering
\includegraphics[width=\columnwidth]{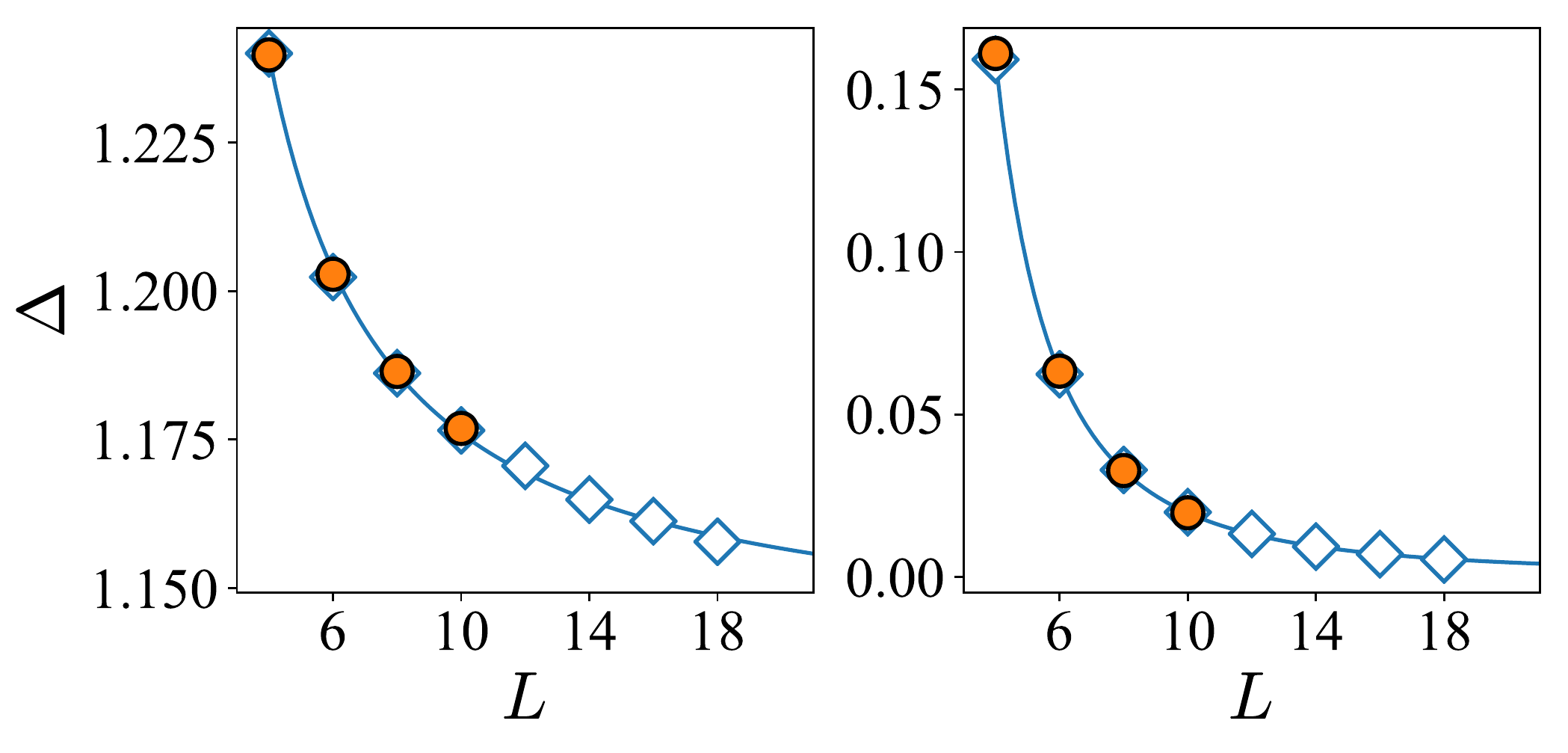}
\caption[largeT_gap]{Infinite temperauture fermion gap as a function of system size for (left) Abelian, $J_x = J_z = 0.25$, $J_z=1$, $K=0$, and (right) non-Abelian, $J_x = J_y = J_z=1$, $K=0.1$, couplings. Both decay quickly towards a fixed values, which from polynomial fits are found to be $\Delta = 1.13(9)$ in the Abelian case and $\Delta = 0.000(6)$ in the non-Abelian case. The fermion gaps obtained from the highest temperature Monte Carlo simulations are displaying (orange circles) on the plots.}
\label{fig:largeT_gap}
\end{figure}

\subsection{Low temperature} 

The ground state of the model lives in the no-vortex sector~\cite{Kitaev2006}. This ground state has a topological degeneracy on a torus, which corresponds to the four different choices of periodic/anti-periodic boundary conditions for the fermions~\cite{Kells2009}. 
Due to the projection procedure we employ, however, some of these topological sectors may have odd fermion parity and so higher ground state energy. 
We take the zero temperature gauge configuration to be an equal weight mixture of the even parity topological sectors, i.e. $\Delta$ is computed as the mean of the fermion gaps of the even parity no-vortex sectors.
In the non-Abelian case this is three of the four sectors, whereas in the Abelian case all four are degenerate. 
We ignore the finite size gapping of the topologically degenerate states, as this vanishes exponentially.

The fermion gap of this zero temperature gauge configuration is plotted as a function of system size for the non-Abelian and Abelian cases in Fig.~\ref{fig:smallT_gap}. Both are decaying to some fixed value, and the non-Abelian case additionally displays oscillations. 
Extrapolating these trends we find the limiting value of the zero temperature fermion gap.

The lowest temperature value of the gap obtained from our Monte Carlo simulation is plotted alongside the curves in Fig.~\ref{fig:smallT_gap}. We find good agreement between the non-Abelian Monte Carlo data and the zero temperature gauge configuration. However, our Abelian data breaks away from it for small systems, $L < 10$. This occurs as a result of the small vortex gap in the Abelian case, recall that in this case the first specific heat peak is at $T \sim 10^{-4}$. Here $1/T$ is in fact comparable to the finite size gapping of the topologically degenerate ground states for $L < 10$ and so the Monte Carlo simulation resolves the degeneracy. Our assumption that the finite size gapping of the topological degeneracy can be neglected fails in these cases. This demonstrates that we must consider systems of size $L = 10$ or larger in the Abelian case.

\subsection{High temperature} 

Large temperatures can be studied by sampling from completely random gauge configurations. Fig.~\ref{fig:largeT_gap} shows the fermion gaps obtained by averaging over 10,000 random $\gras{u}$ patterns as a function of system size $L$. We fit a polynomial decay to the data. In the Abelian case the trend is toward a non-zero gap, whereas in the non-Abelian case the gap vanishes in the large system limit. The Monte Carlo values obtained at the highest temperatures studied are plotted on the graphs and we find excellent agreement.


\begin{thebibliography}{48}
\expandafter\ifx\csname natexlab\endcsname\relax\def\natexlab#1{#1}\fi
\expandafter\ifx\csname bibnamefont\endcsname\relax
  \def\bibnamefont#1{#1}\fi
\expandafter\ifx\csname bibfnamefont\endcsname\relax
  \def\bibfnamefont#1{#1}\fi
\expandafter\ifx\csname citenamefont\endcsname\relax
  \def\citenamefont#1{#1}\fi
\expandafter\ifx\csname url\endcsname\relax
  \def\url#1{\texttt{#1}}\fi
\expandafter\ifx\csname urlprefix\endcsname\relax\def\urlprefix{URL }\fi
\providecommand{\bibinfo}[2]{#2}
\providecommand{\eprint}[2][]{\url{#2}}

\bibitem[{\citenamefont{Savary and Balents}(2017)}]{Savary2016}
\bibinfo{author}{\bibfnamefont{L.}~\bibnamefont{Savary}} \bibnamefont{and}
  \bibinfo{author}{\bibfnamefont{L.}~\bibnamefont{Balents}},
  \bibinfo{journal}{Rept. Prog. Phys.} \textbf{\bibinfo{volume}{80}},
  \bibinfo{pages}{016502} (\bibinfo{year}{2017}).

\bibitem[{\citenamefont{Knolle and Moessner}(2018)}]{knolle2018field}
\bibinfo{author}{\bibfnamefont{J.}~\bibnamefont{Knolle}} \bibnamefont{and}
  \bibinfo{author}{\bibfnamefont{R.}~\bibnamefont{Moessner}},
  \bibinfo{journal}{arXiv preprint arXiv:1804.02037}  (\bibinfo{year}{2018}).

\bibitem[{\citenamefont{Kitaev}(2006)}]{Kitaev2006}
\bibinfo{author}{\bibfnamefont{A.}~\bibnamefont{Kitaev}},
  \bibinfo{journal}{Annals of Physics} \textbf{\bibinfo{volume}{321}},
  \bibinfo{pages}{2} (\bibinfo{year}{2006}).

\bibitem[{\citenamefont{Knolle et~al.}(2014)\citenamefont{Knolle, Kovrizhin,
  Chalker, and Moessner}}]{Knolle2014}
\bibinfo{author}{\bibfnamefont{J.}~\bibnamefont{Knolle}},
  \bibinfo{author}{\bibfnamefont{D.~L.} \bibnamefont{Kovrizhin}},
  \bibinfo{author}{\bibfnamefont{J.~T.} \bibnamefont{Chalker}},
  \bibnamefont{and} \bibinfo{author}{\bibfnamefont{R.}~\bibnamefont{Moessner}},
  \bibinfo{journal}{Phys. Rev. Lett.} \textbf{\bibinfo{volume}{112}},
  \bibinfo{pages}{207203} (\bibinfo{year}{2014}),
  \urlprefix\url{https://link.aps.org/doi/10.1103/PhysRevLett.112.207203}.

\bibitem[{\citenamefont{Knolle et~al.}(2015)\citenamefont{Knolle, Kovrizhin,
  Chalker, and Moessner}}]{Knolle2015a}
\bibinfo{author}{\bibfnamefont{J.}~\bibnamefont{Knolle}},
  \bibinfo{author}{\bibfnamefont{D.~L.} \bibnamefont{Kovrizhin}},
  \bibinfo{author}{\bibfnamefont{J.~T.} \bibnamefont{Chalker}},
  \bibnamefont{and} \bibinfo{author}{\bibfnamefont{R.}~\bibnamefont{Moessner}},
  \bibinfo{journal}{Phys. Rev. B} \textbf{\bibinfo{volume}{92}},
  \bibinfo{pages}{115127} (\bibinfo{year}{2015}),
  \urlprefix\url{https://link.aps.org/doi/10.1103/PhysRevB.92.115127}.

\bibitem[{\citenamefont{Nasu et~al.}(2014)\citenamefont{Nasu, Udagawa, and
  Motome}}]{Nasu2014}
\bibinfo{author}{\bibfnamefont{J.}~\bibnamefont{Nasu}},
  \bibinfo{author}{\bibfnamefont{M.}~\bibnamefont{Udagawa}}, \bibnamefont{and}
  \bibinfo{author}{\bibfnamefont{Y.}~\bibnamefont{Motome}},
  \bibinfo{journal}{Phys. Rev. Lett.} \textbf{\bibinfo{volume}{113}},
  \bibinfo{pages}{197205} (\bibinfo{year}{2014}),
  \urlprefix\url{https://link.aps.org/doi/10.1103/PhysRevLett.113.197205}.

\bibitem[{\citenamefont{Nasu et~al.}(2015)\citenamefont{Nasu, Udagawa, and
  Motome}}]{Nasu2015b}
\bibinfo{author}{\bibfnamefont{J.}~\bibnamefont{Nasu}},
  \bibinfo{author}{\bibfnamefont{M.}~\bibnamefont{Udagawa}}, \bibnamefont{and}
  \bibinfo{author}{\bibfnamefont{Y.}~\bibnamefont{Motome}},
  \bibinfo{journal}{Physical Review B} \textbf{\bibinfo{volume}{92}},
  \bibinfo{pages}{115122} (\bibinfo{year}{2015}).

\bibitem[{\citenamefont{Nasu and Motome}(2015)}]{Nasu2015a}
\bibinfo{author}{\bibfnamefont{J.}~\bibnamefont{Nasu}} \bibnamefont{and}
  \bibinfo{author}{\bibfnamefont{Y.}~\bibnamefont{Motome}},
  \bibinfo{journal}{Phys. Rev. Lett.} \textbf{\bibinfo{volume}{115}},
  \bibinfo{pages}{087203} (\bibinfo{year}{2015}).

\bibitem[{\citenamefont{Jackeli and Khaliullin}(2009)}]{Jackeli2009}
\bibinfo{author}{\bibfnamefont{G.}~\bibnamefont{Jackeli}} \bibnamefont{and}
  \bibinfo{author}{\bibfnamefont{G.}~\bibnamefont{Khaliullin}},
  \bibinfo{journal}{Phys. Rev. Lett.} \textbf{\bibinfo{volume}{102}},
  \bibinfo{pages}{017205} (\bibinfo{year}{2009}),
  \urlprefix\url{https://link.aps.org/doi/10.1103/PhysRevLett.102.017205}.

\bibitem[{\citenamefont{Singh and Gegenwart}(2010)}]{Singh2010}
\bibinfo{author}{\bibfnamefont{Y.}~\bibnamefont{Singh}} \bibnamefont{and}
  \bibinfo{author}{\bibfnamefont{P.}~\bibnamefont{Gegenwart}},
  \bibinfo{journal}{Phys. Rev. B} \textbf{\bibinfo{volume}{82}},
  \bibinfo{pages}{064412} (\bibinfo{year}{2010}),
  \urlprefix\url{https://link.aps.org/doi/10.1103/PhysRevB.82.064412}.

\bibitem[{\citenamefont{Plumb et~al.}(2014)\citenamefont{Plumb, Clancy,
  Sandilands, Shankar, Hu, Burch, Kee, and Kim}}]{Plumb2014}
\bibinfo{author}{\bibfnamefont{K.~W.} \bibnamefont{Plumb}},
  \bibinfo{author}{\bibfnamefont{J.~P.} \bibnamefont{Clancy}},
  \bibinfo{author}{\bibfnamefont{L.~J.} \bibnamefont{Sandilands}},
  \bibinfo{author}{\bibfnamefont{V.~V.} \bibnamefont{Shankar}},
  \bibinfo{author}{\bibfnamefont{Y.~F.} \bibnamefont{Hu}},
  \bibinfo{author}{\bibfnamefont{K.~S.} \bibnamefont{Burch}},
  \bibinfo{author}{\bibfnamefont{H.-Y.} \bibnamefont{Kee}}, \bibnamefont{and}
  \bibinfo{author}{\bibfnamefont{Y.-J.} \bibnamefont{Kim}},
  \bibinfo{journal}{Phys. Rev. B} \textbf{\bibinfo{volume}{90}},
  \bibinfo{pages}{041112} (\bibinfo{year}{2014}),
  \urlprefix\url{https://link.aps.org/doi/10.1103/PhysRevB.90.041112}.

\bibitem[{\citenamefont{Banerjee et~al.}(2016)\citenamefont{Banerjee, Bridges,
  Yan, Aczel, Li, Stone, Granroth, Lumsden, Yiu, Knolle
  et~al.}}]{banerjee2016proximate}
\bibinfo{author}{\bibfnamefont{A.}~\bibnamefont{Banerjee}},
  \bibinfo{author}{\bibfnamefont{C.}~\bibnamefont{Bridges}},
  \bibinfo{author}{\bibfnamefont{J.-Q.} \bibnamefont{Yan}},
  \bibinfo{author}{\bibfnamefont{A.}~\bibnamefont{Aczel}},
  \bibinfo{author}{\bibfnamefont{L.}~\bibnamefont{Li}},
  \bibinfo{author}{\bibfnamefont{M.}~\bibnamefont{Stone}},
  \bibinfo{author}{\bibfnamefont{G.}~\bibnamefont{Granroth}},
  \bibinfo{author}{\bibfnamefont{M.}~\bibnamefont{Lumsden}},
  \bibinfo{author}{\bibfnamefont{Y.}~\bibnamefont{Yiu}},
  \bibinfo{author}{\bibfnamefont{J.}~\bibnamefont{Knolle}},
  \bibnamefont{et~al.}, \bibinfo{journal}{Nature materials}
  \textbf{\bibinfo{volume}{15}}, \bibinfo{pages}{733} (\bibinfo{year}{2016}).

\bibitem[{\citenamefont{Winter et~al.}(2017)\citenamefont{Winter, Tsirlin,
  Daghofer, van~den Brink, Singh, Gegenwart, and Valenti}}]{winter2017models}
\bibinfo{author}{\bibfnamefont{S.~M.} \bibnamefont{Winter}},
  \bibinfo{author}{\bibfnamefont{A.~A.} \bibnamefont{Tsirlin}},
  \bibinfo{author}{\bibfnamefont{M.}~\bibnamefont{Daghofer}},
  \bibinfo{author}{\bibfnamefont{J.}~\bibnamefont{van~den Brink}},
  \bibinfo{author}{\bibfnamefont{Y.}~\bibnamefont{Singh}},
  \bibinfo{author}{\bibfnamefont{P.}~\bibnamefont{Gegenwart}},
  \bibnamefont{and} \bibinfo{author}{\bibfnamefont{R.}~\bibnamefont{Valenti}},
  \bibinfo{journal}{Journal of Physics: Condensed Matter}
  \textbf{\bibinfo{volume}{29}}, \bibinfo{pages}{493002}
  (\bibinfo{year}{2017}).

\bibitem[{\citenamefont{Hermanns et~al.}(2017)\citenamefont{Hermanns, Kimchi,
  and Knolle}}]{hermanns2017physics}
\bibinfo{author}{\bibfnamefont{M.}~\bibnamefont{Hermanns}},
  \bibinfo{author}{\bibfnamefont{I.}~\bibnamefont{Kimchi}}, \bibnamefont{and}
  \bibinfo{author}{\bibfnamefont{J.}~\bibnamefont{Knolle}},
  \bibinfo{journal}{Annual Review of Condensed Matter Physics}
  (\bibinfo{year}{2017}).

\bibitem[{\citenamefont{Hentrich et~al.}(2018)\citenamefont{Hentrich, Wolter,
  Zotos, Brenig, Nowak, Isaeva, Doert, Banerjee, Lampen-Kelley, Mandrus
  et~al.}}]{Hentrich2018unusual}
\bibinfo{author}{\bibfnamefont{R.}~\bibnamefont{Hentrich}},
  \bibinfo{author}{\bibfnamefont{A.~U.~B.} \bibnamefont{Wolter}},
  \bibinfo{author}{\bibfnamefont{X.}~\bibnamefont{Zotos}},
  \bibinfo{author}{\bibfnamefont{W.}~\bibnamefont{Brenig}},
  \bibinfo{author}{\bibfnamefont{D.}~\bibnamefont{Nowak}},
  \bibinfo{author}{\bibfnamefont{A.}~\bibnamefont{Isaeva}},
  \bibinfo{author}{\bibfnamefont{T.}~\bibnamefont{Doert}},
  \bibinfo{author}{\bibfnamefont{A.}~\bibnamefont{Banerjee}},
  \bibinfo{author}{\bibfnamefont{P.}~\bibnamefont{Lampen-Kelley}},
  \bibinfo{author}{\bibfnamefont{D.~G.} \bibnamefont{Mandrus}},
  \bibnamefont{et~al.}, \bibinfo{journal}{Phys. Rev. Lett.}
  \textbf{\bibinfo{volume}{120}}, \bibinfo{pages}{117204}
  (\bibinfo{year}{2018}),
  \urlprefix\url{https://link.aps.org/doi/10.1103/PhysRevLett.120.117204}.

\bibitem[{\citenamefont{Kasahara
  et~al.}(2018{\natexlab{a}})\citenamefont{Kasahara, Sugii, Ohnishi, Shimozawa,
  Yamashita, Kurita, Tanaka, Nasu, Motome, Shibauchi
  et~al.}}]{Kasahara2018unusual}
\bibinfo{author}{\bibfnamefont{Y.}~\bibnamefont{Kasahara}},
  \bibinfo{author}{\bibfnamefont{K.}~\bibnamefont{Sugii}},
  \bibinfo{author}{\bibfnamefont{T.}~\bibnamefont{Ohnishi}},
  \bibinfo{author}{\bibfnamefont{M.}~\bibnamefont{Shimozawa}},
  \bibinfo{author}{\bibfnamefont{M.}~\bibnamefont{Yamashita}},
  \bibinfo{author}{\bibfnamefont{N.}~\bibnamefont{Kurita}},
  \bibinfo{author}{\bibfnamefont{H.}~\bibnamefont{Tanaka}},
  \bibinfo{author}{\bibfnamefont{J.}~\bibnamefont{Nasu}},
  \bibinfo{author}{\bibfnamefont{Y.}~\bibnamefont{Motome}},
  \bibinfo{author}{\bibfnamefont{T.}~\bibnamefont{Shibauchi}},
  \bibnamefont{et~al.}, \bibinfo{journal}{Phys. Rev. Lett.}
  \textbf{\bibinfo{volume}{120}}, \bibinfo{pages}{217205}
  (\bibinfo{year}{2018}{\natexlab{a}}),
  \urlprefix\url{https://link.aps.org/doi/10.1103/PhysRevLett.120.217205}.

\bibitem[{\citenamefont{Kasahara
  et~al.}(2018{\natexlab{b}})\citenamefont{Kasahara, Ohnishi, Mizukami, Tanaka,
  Ma, Sugii, Kurita, Tanaka, Nasu, Motome et~al.}}]{kasahara2018majorana}
\bibinfo{author}{\bibfnamefont{Y.}~\bibnamefont{Kasahara}},
  \bibinfo{author}{\bibfnamefont{T.}~\bibnamefont{Ohnishi}},
  \bibinfo{author}{\bibfnamefont{Y.}~\bibnamefont{Mizukami}},
  \bibinfo{author}{\bibfnamefont{O.}~\bibnamefont{Tanaka}},
  \bibinfo{author}{\bibfnamefont{S.}~\bibnamefont{Ma}},
  \bibinfo{author}{\bibfnamefont{K.}~\bibnamefont{Sugii}},
  \bibinfo{author}{\bibfnamefont{N.}~\bibnamefont{Kurita}},
  \bibinfo{author}{\bibfnamefont{H.}~\bibnamefont{Tanaka}},
  \bibinfo{author}{\bibfnamefont{J.}~\bibnamefont{Nasu}},
  \bibinfo{author}{\bibfnamefont{Y.}~\bibnamefont{Motome}},
  \bibnamefont{et~al.}, \bibinfo{journal}{arXiv preprint arXiv:1805.05022}
  (\bibinfo{year}{2018}{\natexlab{b}}).

\bibitem[{\citenamefont{Sears et~al.}(2017)\citenamefont{Sears, Zhao, Xu, Lynn,
  and Kim}}]{Sears2017phase}
\bibinfo{author}{\bibfnamefont{J.~A.} \bibnamefont{Sears}},
  \bibinfo{author}{\bibfnamefont{Y.}~\bibnamefont{Zhao}},
  \bibinfo{author}{\bibfnamefont{Z.}~\bibnamefont{Xu}},
  \bibinfo{author}{\bibfnamefont{J.~W.} \bibnamefont{Lynn}}, \bibnamefont{and}
  \bibinfo{author}{\bibfnamefont{Y.-J.} \bibnamefont{Kim}},
  \bibinfo{journal}{Phys. Rev. B} \textbf{\bibinfo{volume}{95}},
  \bibinfo{pages}{180411} (\bibinfo{year}{2017}),
  \urlprefix\url{https://link.aps.org/doi/10.1103/PhysRevB.95.180411}.

\bibitem[{\citenamefont{Baek et~al.}(2017)\citenamefont{Baek, Do, Choi, Kwon,
  Wolter, Nishimoto, van~den Brink, and B\"uchner}}]{Baek2017evidence}
\bibinfo{author}{\bibfnamefont{S.-H.} \bibnamefont{Baek}},
  \bibinfo{author}{\bibfnamefont{S.-H.} \bibnamefont{Do}},
  \bibinfo{author}{\bibfnamefont{K.-Y.} \bibnamefont{Choi}},
  \bibinfo{author}{\bibfnamefont{Y.~S.} \bibnamefont{Kwon}},
  \bibinfo{author}{\bibfnamefont{A.~U.~B.} \bibnamefont{Wolter}},
  \bibinfo{author}{\bibfnamefont{S.}~\bibnamefont{Nishimoto}},
  \bibinfo{author}{\bibfnamefont{J.}~\bibnamefont{van~den Brink}},
  \bibnamefont{and}
  \bibinfo{author}{\bibfnamefont{B.}~\bibnamefont{B\"uchner}},
  \bibinfo{journal}{Phys. Rev. Lett.} \textbf{\bibinfo{volume}{119}},
  \bibinfo{pages}{037201} (\bibinfo{year}{2017}),
  \urlprefix\url{https://link.aps.org/doi/10.1103/PhysRevLett.119.037201}.

\bibitem[{\citenamefont{Banerjee et~al.}(2018)\citenamefont{Banerjee,
  Lampen-Kelley, Knolle, Balz, Aczel, Winn, Liu, Pajerowski, Yan, Bridges
  et~al.}}]{banerjee2018excitations}
\bibinfo{author}{\bibfnamefont{A.}~\bibnamefont{Banerjee}},
  \bibinfo{author}{\bibfnamefont{P.}~\bibnamefont{Lampen-Kelley}},
  \bibinfo{author}{\bibfnamefont{J.}~\bibnamefont{Knolle}},
  \bibinfo{author}{\bibfnamefont{C.}~\bibnamefont{Balz}},
  \bibinfo{author}{\bibfnamefont{A.~A.} \bibnamefont{Aczel}},
  \bibinfo{author}{\bibfnamefont{B.}~\bibnamefont{Winn}},
  \bibinfo{author}{\bibfnamefont{Y.}~\bibnamefont{Liu}},
  \bibinfo{author}{\bibfnamefont{D.}~\bibnamefont{Pajerowski}},
  \bibinfo{author}{\bibfnamefont{J.}~\bibnamefont{Yan}},
  \bibinfo{author}{\bibfnamefont{C.~A.} \bibnamefont{Bridges}},
  \bibnamefont{et~al.}, \bibinfo{journal}{npj Quantum Materials}
  \textbf{\bibinfo{volume}{3}}, \bibinfo{pages}{8} (\bibinfo{year}{2018}).

\bibitem[{\citenamefont{Lahtinen}(2011)}]{Lahtinen:2011gm}
\bibinfo{author}{\bibfnamefont{V.}~\bibnamefont{Lahtinen}},
  \bibinfo{journal}{New Journal of Physics} \textbf{\bibinfo{volume}{13}},
  \bibinfo{pages}{075009} (\bibinfo{year}{2011}).

\bibitem[{\citenamefont{Altland and Zirnbauer}(1997)}]{Altland1997}
\bibinfo{author}{\bibfnamefont{A.}~\bibnamefont{Altland}} \bibnamefont{and}
  \bibinfo{author}{\bibfnamefont{M.~R.} \bibnamefont{Zirnbauer}},
  \bibinfo{journal}{Phys. Rev. B} \textbf{\bibinfo{volume}{55}},
  \bibinfo{pages}{1142} (\bibinfo{year}{1997}),
  \urlprefix\url{https://link.aps.org/doi/10.1103/PhysRevB.55.1142}.

\bibitem[{\citenamefont{Chalker et~al.}(2001)\citenamefont{Chalker, Read,
  Kagalovsky, Horovitz, Avishai, and Ludwig}}]{Chalker2001}
\bibinfo{author}{\bibfnamefont{J.~T.} \bibnamefont{Chalker}},
  \bibinfo{author}{\bibfnamefont{N.}~\bibnamefont{Read}},
  \bibinfo{author}{\bibfnamefont{V.}~\bibnamefont{Kagalovsky}},
  \bibinfo{author}{\bibfnamefont{B.}~\bibnamefont{Horovitz}},
  \bibinfo{author}{\bibfnamefont{Y.}~\bibnamefont{Avishai}}, \bibnamefont{and}
  \bibinfo{author}{\bibfnamefont{A.~W.~W.} \bibnamefont{Ludwig}},
  \bibinfo{journal}{Phys. Rev. B} \textbf{\bibinfo{volume}{65}},
  \bibinfo{pages}{012506} (\bibinfo{year}{2001}),
  \urlprefix\url{https://link.aps.org/doi/10.1103/PhysRevB.65.012506}.

\bibitem[{\citenamefont{Mildenberger et~al.}(2007)\citenamefont{Mildenberger,
  Evers, Mirlin, and Chalker}}]{Mildenberger2007}
\bibinfo{author}{\bibfnamefont{A.}~\bibnamefont{Mildenberger}},
  \bibinfo{author}{\bibfnamefont{F.}~\bibnamefont{Evers}},
  \bibinfo{author}{\bibfnamefont{A.~D.} \bibnamefont{Mirlin}},
  \bibnamefont{and} \bibinfo{author}{\bibfnamefont{J.~T.}
  \bibnamefont{Chalker}}, \bibinfo{journal}{Phys. Rev. B}
  \textbf{\bibinfo{volume}{75}}, \bibinfo{pages}{245321}
  (\bibinfo{year}{2007}),
  \urlprefix\url{https://link.aps.org/doi/10.1103/PhysRevB.75.245321}.

\bibitem[{\citenamefont{Laumann et~al.}(2012)\citenamefont{Laumann, Ludwig,
  Huse, and Trebst}}]{Laumann2012}
\bibinfo{author}{\bibfnamefont{C.~R.} \bibnamefont{Laumann}},
  \bibinfo{author}{\bibfnamefont{A.~W.~W.} \bibnamefont{Ludwig}},
  \bibinfo{author}{\bibfnamefont{D.~A.} \bibnamefont{Huse}}, \bibnamefont{and}
  \bibinfo{author}{\bibfnamefont{S.}~\bibnamefont{Trebst}},
  \bibinfo{journal}{Physical Review B} \textbf{\bibinfo{volume}{85}},
  \bibinfo{pages}{1352} (\bibinfo{year}{2012}).

\bibitem[{\citenamefont{Beenakker}(2013)}]{Beenakker2013}
\bibinfo{author}{\bibfnamefont{C.~W.~J.} \bibnamefont{Beenakker}},
  \bibinfo{journal}{Annual Review of Condensed Matter Physics}
  \textbf{\bibinfo{volume}{4}}, \bibinfo{pages}{113} (\bibinfo{year}{2013}).

\bibitem[{\citenamefont{Metavitsiadis et~al.}(2017)\citenamefont{Metavitsiadis,
  Pidatella, and Brenig}}]{Metavitsiadis2017}
\bibinfo{author}{\bibfnamefont{A.}~\bibnamefont{Metavitsiadis}},
  \bibinfo{author}{\bibfnamefont{A.}~\bibnamefont{Pidatella}},
  \bibnamefont{and} \bibinfo{author}{\bibfnamefont{W.}~\bibnamefont{Brenig}},
  \bibinfo{journal}{Physical Review B} \textbf{\bibinfo{volume}{96}},
  \bibinfo{pages}{205121} (\bibinfo{year}{2017}).

\bibitem[{\citenamefont{Evers and Mirlin}(2008)}]{Evers2008}
\bibinfo{author}{\bibfnamefont{F.}~\bibnamefont{Evers}} \bibnamefont{and}
  \bibinfo{author}{\bibfnamefont{A.~D.} \bibnamefont{Mirlin}},
  \bibinfo{journal}{Rev. Mod. Phys.} \textbf{\bibinfo{volume}{80}},
  \bibinfo{pages}{1355} (\bibinfo{year}{2008}),
  \urlprefix\url{https://link.aps.org/doi/10.1103/RevModPhys.80.1355}.

\bibitem[{\citenamefont{Lahtinen and Pachos}(2009)}]{Lahtinen2009}
\bibinfo{author}{\bibfnamefont{V.}~\bibnamefont{Lahtinen}} \bibnamefont{and}
  \bibinfo{author}{\bibfnamefont{J.~K.} \bibnamefont{Pachos}},
  \bibinfo{journal}{New J. Phys.} \textbf{\bibinfo{volume}{11}},
  \bibinfo{pages}{093027} (\bibinfo{year}{2009}).

\bibitem[{\citenamefont{Pedrocchi et~al.}(2011)\citenamefont{Pedrocchi, Chesi,
  and Loss}}]{Pedrocchi2011}
\bibinfo{author}{\bibfnamefont{F.~L.} \bibnamefont{Pedrocchi}},
  \bibinfo{author}{\bibfnamefont{S.}~\bibnamefont{Chesi}}, \bibnamefont{and}
  \bibinfo{author}{\bibfnamefont{D.}~\bibnamefont{Loss}},
  \bibinfo{journal}{Physical Review B} \textbf{\bibinfo{volume}{84}},
  \bibinfo{pages}{165414} (\bibinfo{year}{2011}).

\bibitem[{\citenamefont{Self}(in preparation)}]{chrisUpcoming}
\bibinfo{author}{\bibfnamefont{C.~N.} \bibnamefont{Self}} (\bibinfo{year}{in
  preparation}).

\bibitem[{\citenamefont{Efron}(1979)}]{Efron1979}
\bibinfo{author}{\bibfnamefont{B.}~\bibnamefont{Efron}}, \bibinfo{journal}{SIAM
  Review} \textbf{\bibinfo{volume}{21}}, \bibinfo{pages}{460}
  (\bibinfo{year}{1979}).

\bibitem[{\citenamefont{Iblisdir et~al.}(2009)\citenamefont{Iblisdir,
  P\'erez-Garc\'{\i}a, Aguado, and Pachos}}]{Iblisdir2009scaling}
\bibinfo{author}{\bibfnamefont{S.}~\bibnamefont{Iblisdir}},
  \bibinfo{author}{\bibfnamefont{D.}~\bibnamefont{P\'erez-Garc\'{\i}a}},
  \bibinfo{author}{\bibfnamefont{M.}~\bibnamefont{Aguado}}, \bibnamefont{and}
  \bibinfo{author}{\bibfnamefont{J.}~\bibnamefont{Pachos}},
  \bibinfo{journal}{Phys. Rev. B} \textbf{\bibinfo{volume}{79}},
  \bibinfo{pages}{134303} (\bibinfo{year}{2009}),
  \urlprefix\url{https://link.aps.org/doi/10.1103/PhysRevB.79.134303}.

\bibitem[{\citenamefont{Iblisdir et~al.}(2010)\citenamefont{Iblisdir,
  Perez-Garcia, Aguado, and Pachos}}]{iblisdir2010thermal}
\bibinfo{author}{\bibfnamefont{S.}~\bibnamefont{Iblisdir}},
  \bibinfo{author}{\bibfnamefont{D.}~\bibnamefont{Perez-Garcia}},
  \bibinfo{author}{\bibfnamefont{M.}~\bibnamefont{Aguado}}, \bibnamefont{and}
  \bibinfo{author}{\bibfnamefont{J.}~\bibnamefont{Pachos}},
  \bibinfo{journal}{Nuclear Physics B} \textbf{\bibinfo{volume}{829}},
  \bibinfo{pages}{401} (\bibinfo{year}{2010}).

\bibitem[{\citenamefont{Lahtinen et~al.}(2012)\citenamefont{Lahtinen, Ludwig,
  Pachos, and Trebst}}]{Lahtinen2012}
\bibinfo{author}{\bibfnamefont{V.}~\bibnamefont{Lahtinen}},
  \bibinfo{author}{\bibfnamefont{A.~W.~W.} \bibnamefont{Ludwig}},
  \bibinfo{author}{\bibfnamefont{J.~K.} \bibnamefont{Pachos}},
  \bibnamefont{and} \bibinfo{author}{\bibfnamefont{S.}~\bibnamefont{Trebst}},
  \bibinfo{journal}{Physical Review B} \textbf{\bibinfo{volume}{86}},
  \bibinfo{pages}{075115} (\bibinfo{year}{2012}).

\bibitem[{\citenamefont{Lahtinen et~al.}(2014)\citenamefont{Lahtinen, Ludwig,
  and Trebst}}]{Lahtinen2014}
\bibinfo{author}{\bibfnamefont{V.}~\bibnamefont{Lahtinen}},
  \bibinfo{author}{\bibfnamefont{A.~W.~W.} \bibnamefont{Ludwig}},
  \bibnamefont{and} \bibinfo{author}{\bibfnamefont{S.}~\bibnamefont{Trebst}},
  \bibinfo{journal}{Physical Review B} \textbf{\bibinfo{volume}{89}},
  \bibinfo{pages}{085121} (\bibinfo{year}{2014}).

\bibitem[{\citenamefont{Bauer et~al.}(2013)\citenamefont{Bauer, Lutchyn,
  Hastings, and Troyer}}]{Bauer2013}
\bibinfo{author}{\bibfnamefont{B.}~\bibnamefont{Bauer}},
  \bibinfo{author}{\bibfnamefont{R.~M.} \bibnamefont{Lutchyn}},
  \bibinfo{author}{\bibfnamefont{M.~B.} \bibnamefont{Hastings}},
  \bibnamefont{and} \bibinfo{author}{\bibfnamefont{M.}~\bibnamefont{Troyer}},
  \bibinfo{journal}{Phys. Rev. B} \textbf{\bibinfo{volume}{87}},
  \bibinfo{pages}{014503} (\bibinfo{year}{2013}),
  \urlprefix\url{https://link.aps.org/doi/10.1103/PhysRevB.87.014503}.

\bibitem[{\citenamefont{Bocquet et~al.}(2000)\citenamefont{Bocquet, Serban, and
  Zirnbauer}}]{Bocquet:1999bb}
\bibinfo{author}{\bibfnamefont{M.}~\bibnamefont{Bocquet}},
  \bibinfo{author}{\bibfnamefont{D.}~\bibnamefont{Serban}}, \bibnamefont{and}
  \bibinfo{author}{\bibfnamefont{M.~R.} \bibnamefont{Zirnbauer}},
  \bibinfo{journal}{Nuclear Physics B} \textbf{\bibinfo{volume}{578}},
  \bibinfo{pages}{628} (\bibinfo{year}{2000}).

\bibitem[{\citenamefont{Wigner}(1993)}]{wigner1993class}
\bibinfo{author}{\bibfnamefont{E.~P.} \bibnamefont{Wigner}}, in
  \emph{\bibinfo{booktitle}{The Collected Works of Eugene Paul Wigner}}
  (\bibinfo{publisher}{Springer}, \bibinfo{year}{1993}), pp.
  \bibinfo{pages}{409--440}.

\bibitem[{\citenamefont{Fal'ko and Efetov}(2007)}]{Falko:2007bl}
\bibinfo{author}{\bibfnamefont{V.~I.} \bibnamefont{Fal'ko}} \bibnamefont{and}
  \bibinfo{author}{\bibfnamefont{K.~B.} \bibnamefont{Efetov}},
  \bibinfo{journal}{Europhysics Letters (EPL)} \textbf{\bibinfo{volume}{32}},
  \bibinfo{pages}{627} (\bibinfo{year}{2007}).

\bibitem[{\citenamefont{Sandilands et~al.}(2015)\citenamefont{Sandilands, Tian,
  Plumb, Kim, and Burch}}]{Sandilands2015scattering}
\bibinfo{author}{\bibfnamefont{L.~J.} \bibnamefont{Sandilands}},
  \bibinfo{author}{\bibfnamefont{Y.}~\bibnamefont{Tian}},
  \bibinfo{author}{\bibfnamefont{K.~W.} \bibnamefont{Plumb}},
  \bibinfo{author}{\bibfnamefont{Y.-J.} \bibnamefont{Kim}}, \bibnamefont{and}
  \bibinfo{author}{\bibfnamefont{K.~S.} \bibnamefont{Burch}},
  \bibinfo{journal}{Phys. Rev. Lett.} \textbf{\bibinfo{volume}{114}},
  \bibinfo{pages}{147201} (\bibinfo{year}{2015}),
  \urlprefix\url{https://link.aps.org/doi/10.1103/PhysRevLett.114.147201}.

\bibitem[{\citenamefont{Perreault et~al.}(2016)\citenamefont{Perreault, Knolle,
  Perkins, and Burnell}}]{Brent2016}
\bibinfo{author}{\bibfnamefont{B.}~\bibnamefont{Perreault}},
  \bibinfo{author}{\bibfnamefont{J.}~\bibnamefont{Knolle}},
  \bibinfo{author}{\bibfnamefont{N.~B.} \bibnamefont{Perkins}},
  \bibnamefont{and} \bibinfo{author}{\bibfnamefont{F.~J.}
  \bibnamefont{Burnell}}, \bibinfo{journal}{Phys. Rev. B}
  \textbf{\bibinfo{volume}{94}}, \bibinfo{pages}{104427}
  (\bibinfo{year}{2016}),
  \urlprefix\url{https://link.aps.org/doi/10.1103/PhysRevB.94.104427}.

\bibitem[{\citenamefont{Smith et~al.}(2017)\citenamefont{Smith, Knolle,
  Kovrizhin, and Moessner}}]{Smith2017}
\bibinfo{author}{\bibfnamefont{A.}~\bibnamefont{Smith}},
  \bibinfo{author}{\bibfnamefont{J.}~\bibnamefont{Knolle}},
  \bibinfo{author}{\bibfnamefont{D.~L.} \bibnamefont{Kovrizhin}},
  \bibnamefont{and} \bibinfo{author}{\bibfnamefont{R.}~\bibnamefont{Moessner}},
  \bibinfo{journal}{Phys. Rev. Lett.} \textbf{\bibinfo{volume}{118}},
  \bibinfo{pages}{266601} (\bibinfo{year}{2017}),
  \urlprefix\url{https://link.aps.org/doi/10.1103/PhysRevLett.118.266601}.

\bibitem[{\citenamefont{Prosko et~al.}(2017)\citenamefont{Prosko, Lee, and
  Maciejko}}]{Prosko2017}
\bibinfo{author}{\bibfnamefont{C.}~\bibnamefont{Prosko}},
  \bibinfo{author}{\bibfnamefont{S.-P.} \bibnamefont{Lee}}, \bibnamefont{and}
  \bibinfo{author}{\bibfnamefont{J.}~\bibnamefont{Maciejko}},
  \bibinfo{journal}{Phys. Rev. B} \textbf{\bibinfo{volume}{96}},
  \bibinfo{pages}{205104} (\bibinfo{year}{2017}),
  \urlprefix\url{https://link.aps.org/doi/10.1103/PhysRevB.96.205104}.

\bibitem[{\citenamefont{Smith et~al.}(2018)\citenamefont{Smith, Knolle,
  Moessner, and Kovrizhin}}]{smith2018dynamical}
\bibinfo{author}{\bibfnamefont{A.}~\bibnamefont{Smith}},
  \bibinfo{author}{\bibfnamefont{J.}~\bibnamefont{Knolle}},
  \bibinfo{author}{\bibfnamefont{R.}~\bibnamefont{Moessner}}, \bibnamefont{and}
  \bibinfo{author}{\bibfnamefont{D.~L.} \bibnamefont{Kovrizhin}},
  \bibinfo{journal}{arXiv preprint arXiv:1803.06574}  (\bibinfo{year}{2018}).

\bibitem[{\citenamefont{Earl and Deem}(2005)}]{Earl2005}
\bibinfo{author}{\bibfnamefont{D.~J.} \bibnamefont{Earl}} \bibnamefont{and}
  \bibinfo{author}{\bibfnamefont{M.~W.} \bibnamefont{Deem}},
  \bibinfo{journal}{Phys. Chem. Chem. Phys.} \textbf{\bibinfo{volume}{7}},
  \bibinfo{pages}{3910} (\bibinfo{year}{2005}).

\bibitem[{\citenamefont{Geyer}(1992)}]{Geyer1992}
\bibinfo{author}{\bibfnamefont{C.~J.} \bibnamefont{Geyer}},
  \bibinfo{journal}{Statistical Science} \textbf{\bibinfo{volume}{7}},
  \bibinfo{pages}{473} (\bibinfo{year}{1992}).

\bibitem[{\citenamefont{Kells et~al.}(2009)\citenamefont{Kells, Slingerland,
  and Vala}}]{Kells2009}
\bibinfo{author}{\bibfnamefont{G.}~\bibnamefont{Kells}},
  \bibinfo{author}{\bibfnamefont{J.~K.} \bibnamefont{Slingerland}},
  \bibnamefont{and} \bibinfo{author}{\bibfnamefont{J.}~\bibnamefont{Vala}},
  \bibinfo{journal}{Physical Review B} \textbf{\bibinfo{volume}{80}},
  \bibinfo{pages}{125415} (\bibinfo{year}{2009}).

\end{thebibliography}
\end{document}